\documentclass[twocolumn]{aastex631}

\newcommand\JunoCam{{\it JunoCam}}
\newcommand\Juno{{\it Juno}}

\usepackage{amsmath}
\usepackage{amssymb}
\usepackage{gensymb}

\begin{document}

\title{Jovian Vortex Hunter: a citizen science project to study Jupiter's vortices}

\author[0000-0002-6794-7587]{Ramanakumar Sankar}
\affiliation{University of California, \\
501 Campbell Hall, \\
Berkeley, CA 94720, USA}

\author[0000-0002-3669-0539]{Shawn Brueshaber}
\affiliation{Michigan Technological University, \\
Mechanical Engineering-Engineering Mechanics,\\
1400 Houghton, MI 49331 USA}

\author[0000-0002-1067-8558]{Lucy Fortson}
\affiliation{
School of Physics and Astronomy,\\ University of Minnesota, \\
116 Church St SE, \\
Minneapolis, MN 55455, USA}
\affiliation{
Minnesota Institute for Astrophysics,\\ University of Minnesota, \\
116 Church St SE, \\
Minneapolis, MN 55455, USA}

\author{Candice Hansen-Koharcheck}
\affiliation{Planetary Science Institute}

\author[0000-0001-5578-359X]{Chris Lintott}
\affiliation{
Department of Physics,\\
University of Oxford,\\
Denys Wilkinson Building,\\
Keble Road,\\
Oxford, OX1 3RH, United Kingdom}

\author[0000-0002-6016-300X]{Kameswara Mantha}
\affiliation{
School of Physics and Astronomy,\\ University of Minnesota, \\
116 Church St SE, \\
Minneapolis, MN 55455, USA}
\affiliation{
Minnesota Institute for Astrophysics,\\ University of Minnesota, \\
116 Church St SE, \\
Minneapolis, MN 55455, USA}

\author{Cooper Nesmith}
\affiliation{
School of Physics and Astronomy,\\ University of Minnesota, \\
116 Church St SE, \\
Minneapolis, MN 55455, USA}

\author[0000-0001-7871-2823]{Glenn S. Orton}
\affiliation{MS 183-501 \\
Jet Propulsion Laboratory \\
California Institute of Technology \\
4800 Oak Grove Drive \\
Pasadena, CA 91109}



\begin{abstract}
The Jovian atmosphere contains a wide diversity of vortices, which have a large range of sizes, colors and forms in different dynamical regimes. The formation processes for these vortices is poorly understood, and aside from a few known, long-lived ovals, such as the Great Red Spot, and Oval BA, vortex stability and their temporal evolution are currently largely unknown. In this study, we use JunoCam data and a citizen-science project on Zooniverse to derive a catalog of vortices, some with repeated observations, through May 2018 to Sep 2021, and analyze their associated properties, such as size, location and color. We find that different colored vortices (binned as white, red, brown and dark), follow vastly different distributions in terms of their sizes and where they are found on the planet. We employ a simplified stability criterion using these vortices as a proxy, to derive a minimum Rossby deformation length for the planet of $\sim1800$ km. We find that this value of $L_d$ is largely constant throughout the atmosphere, and does not have an appreciable meridional gradient.
\end{abstract}

\keywords{Jupiter}


\section{Introduction} \label{sec:intro}

The Jovian atmosphere presents a diverse set of cloud features, varying in size and color. These different cloud types are governed by the interplay between the underlying atmospheric dynamics and condensation chemistry. On Earth, this interplay is generally better understood because the relation between cloud habit and cloud types are empirically determined using atmospheric sounding instruments and in-situ aircraft measurements \citep{PruppacherKlettbookCh10,Houze2014}. Since this is not possible on Jupiter, we need to extrapolate from our current theories of cloud formation to interpret the formation of these features. Furthermore, since the underlying dynamics is obscured by the upper cloud deck, it is difficult to determine the nature of the deeper atmosphere which is where most of the processes that drive cloud formation are present\citep{Fletcher2020}, specifically at or below the water cloud layer ($\sim100-150$ km below the upper cloud deck where these features are visible). Many of these deeper processes emerge through the formation of dynamical instabilities with significant vertical motion. Long-term interplay between convective outbreaks and the background atmosphere leads to a spectrum of different effects on the cloud layer, from disrupting cloud bands \citep{Fletcher2017,SanchezLavega2017,Sankar2022} to formation of vortical structures \citep{Inurrigarro2020,Hueso2022}. Particularly, several classes of clouds are driven by the formation and evolution of dynamical instabilities, such as convective outbursts and the formation of eddies through shear instabilities \citep{Palotai2022}.  The differences between the resulting structures depends strongly on both the type of the instability (e.g., strength and nature of convective outbreak, barotropic vs baroclinic instabilities, etc.), as well as on the underlying structure of the atmosphere, given by the vertical stratification, vertical wind shear, etc. \citep{GarciaMelendo2005}. Thus, using the observed discrete features (such as vortices) on the visible cloud layer as a proxy, we can infer properties of the deeper atmosphere. 

Primarily, we need a better understanding of the dynamics and structure of the `weather layer' on Jupiter (i.e., the vertical extent of the atmosphere where most of the cloud formation occurs). Several studies have previously investigated individual morphologies related to such instabilities \citep[e.g.,][]{GarciaMelendo2005, Choi2013, SanchezLavega2017}, and while some studies have investigated a sample of hand-counted vortices \citep[e.g., including only vortices larger than 1000 km,][]{Trammell2014}, a comprehensive survey of the array of vortical phenomena across the planet is missing. Numerical and theoretical studies have placed constraints on different types of eddy formation \citep[e.g.,][]{GarciaMelendo2005,Showman2007,Brueshaber2019,Inurrigarro2022}, along with specific criteria on how and where instabilities form \citep[e.g.,][]{Rayleigh1880, Kuo1949, CharneyStern, Dowling1995}. However, these models have failed to generalize across the planets mostly due to the large diversity of the types of instabilities that occur ubiquitously throughout the Jovian atmosphere due to a lack of observation of the deep weather layer. While Juno's Microwave Radiometer  provides some observations of these large ovals \citep{Bolton2021} and discrete instabilities \citep{Fletcher2020JGR}, a comprehensive, global study is still incomplete.
There is a fundamental gap in knowledge about the structure of the deeper atmosphere and the specific triggers for these instabilities, which makes it difficult to draw generalized conclusions about the role that atmospheric processes in the deeper atmosphere play in jovian atmospheric dynamics.

In an effort to address these issues, we instead choose a data-driven approach to study the distribution of cloud morphologies associated with discrete vortical phenomena on Jupiter. We use data from the \JunoCam{} instrument onboard the \Juno{} spacecraft to identify vortices from perijoves 13 through 36 (i.e., observations from May 2018 through September 2021) and study their corresponding distribution in an effort to constrain the properties of the deep atmosphere, particularly relating to the stability of the deep weather layer. To do so requires a catalog of vortices and their associated properties, which is a monumental effort for a single research team. Therefore, in this study, we employ the citizen science methodology to gather the distribution of physical parameters of Jovian vortices, wherein we employ participation from volunteers from across the globe to classify and annotate vortices in the \JunoCam{} images.
The benefits of citizen science in classification and annotation tasks are two-fold: first, citizen scientists can be quickly trained on a small sample of pre-selected data, which is generally difficult to do with machine models. Secondly, classifications from multiple volunteers can be used for the same object in order to build a consensus value and quantify the error in the classification \citep{Lintott2008}. Recently, citizen science has seen extensive growth in astronomy and planetary science, with several projects focused on data gathering, e.g., amateur observations of planets \citep{Hueso2010} or impact flashes \citep{Sankar2020}, and in classifications or annotation, e.g., identification of surface features on Mars \citep{Aye2019}. For a broad review of citizen science efforts in astronomy and space science, see \citet{Marshall2015} and \citet{fortson2021green}. Citizen science has also been vital in the engagement of the community in scientific research, leading to increased science learning \citep{Masters2016} and in the serendipitous discovery of unknown-unknowns -- objects in the dataset that contain previously unknown scientific phenomena \citep[see][and references therein]{Trouille2019}, which lead to interesting new breakthroughs. We also note, there is a nice synergy between \JunoCam{} as an outreach camera and the development of a vortex catalog by citizen scientist volunteers.

Consequently, to build our catalog of vortices for this study, we built and deployed the Jovian Vortex Hunter (JVH) project on the Zooniverse.org citizen science platform. Zooniverse is an online citizen science platform where project teams can host their data and build simple interfaces for volunteers to interact with the data. Zooniverse features a wide diversity of project domains from digital humanities to astrophysics, and boasts more than 2.5 million registered volunteers who, combined, perform several hundred thousand classifications every month. 
To deploy the JVH project, we used the Zooniverse Project Builder\footnote{https://www.zooniverse.org/lab} to build a simple interface where volunteers were asked to choose
images that contained vortices, and annotate their shapes and colors. We subsequently processed and aggregated the volunteer responses to derive the property distributions for the aforementioned perijoves. 

In this study, we describe our first results from the Zooniverse project, our data collection and aggregation methods, and a discussion of the resulting observed vortex properties. In Section~\ref{sec:data}, we discuss the \JunoCam{} data pipeline and our method for creating the volunteer-facing imagery. In Section~\ref{sec:methods}, we describe the JVH project and the aggregation methodology. In Section~\ref{sec:results}, we showcase the vortex property distributions determined from the aggregated volunteer responses. In Section~\ref{sec:discussion}, we discuss these results in the context of global jovian atmospheric dynamics and conclude in Section~\ref{sec:conclusions}.








\section{Data} \label{sec:data}
The data for this project are obtained from the \JunoCam{} instrument on \Juno{}. 

\subsection{\JunoCam{} data}
\JunoCam{} is a push-broom imager on \Juno{} featuring 3 color channels (approximately blue, green and red), together with a ``methane" filter centered on a CH4 absorption band at 800-900 nm that was not used in this study. The camera was added to the mission designed for outreach purposes \citep{Hansen2017} but has nonetheless been used in scientific studies of the dynamics of the jovian atmosphere \citep[e.g.,][]{SanchezLavega2018}, as well as properties of waves \citep{Orton2020} and polar cyclones \citep{Adriani2018,TabatabaVakili2020}. These images are taken roughly within 2 hours of a \Juno{} perijove pass with the lowest resolution of the images typically being during approach and departure over the north and south pole, respectively, and the highest spatial resolution being over the mid- and low-latitude regions. At its closest distance to Jupiter, \JunoCam{} can achieve pixel resolutions as high at 5 km/pixel, making it incredibly useful to identify and study the structures of small-scale cloud features. However, there are several challenges with \JunoCam{}: firstly, due to the lack of a dedicated in-flight calibration pipeline, \JunoCam{} images are generally photometrically uncalibrated, making these images difficult to use in atmospheric retrieval studies. Secondly, due to the short image capture cadence during a perijove (i.e., the features are separated by at most an hour, which is insufficient to detect motion in the clouds), it is difficult to derive wind speed measurements for dynamical studies \citep{SanchezLavega2018}. In this project, we bypass both these issues, since the methodology used in this study does not require strong photometric calibration of the data (only color balancing across the perijoves to aid volunteers' interpretation of vortex colors), and the wind speed measurements are not needed to identify and characterize vortices. Even so, the data require additional processing to project the images and adjust for illumination, as detailed below. For this project, we use data from perijoves 13 through 36. Before perijove 13, we found that many of the images from JunoCam were close to the planet's limb (from Juno's perspective), which led to very narrow (longitudinally) footprints on the planet. As such, it was more challenging to create global maps easily for the earlier perijoves.

\subsection{\JunoCam{} data processing}
The raw \JunoCam{} images are obtained from the Mission Juno website and consist of a series of frames, with three ``framelets" corresponding to the three filters (Blue, Green and Red). Each full image has about 14 frames, and 42 framelets. We use SPICE kernels \citep{Acton1996} from the NAIF server\footnote{\url{https://naif.jpl.nasa.gov/pub/naif/JUNO/kernels/}} to determine the location of each pixel that intersects with Jupiter's reference ellipsoid and determine the planetocentric coordinate of the pixel. Finally, we map the complete image  to a cylindrical planetocentic projection with a resolution of 45 pixels per degree.

\subsubsection{Color correction, lighting correction and mosaicing}
\begin{figure}
    \centering
    \includegraphics[width=\columnwidth]{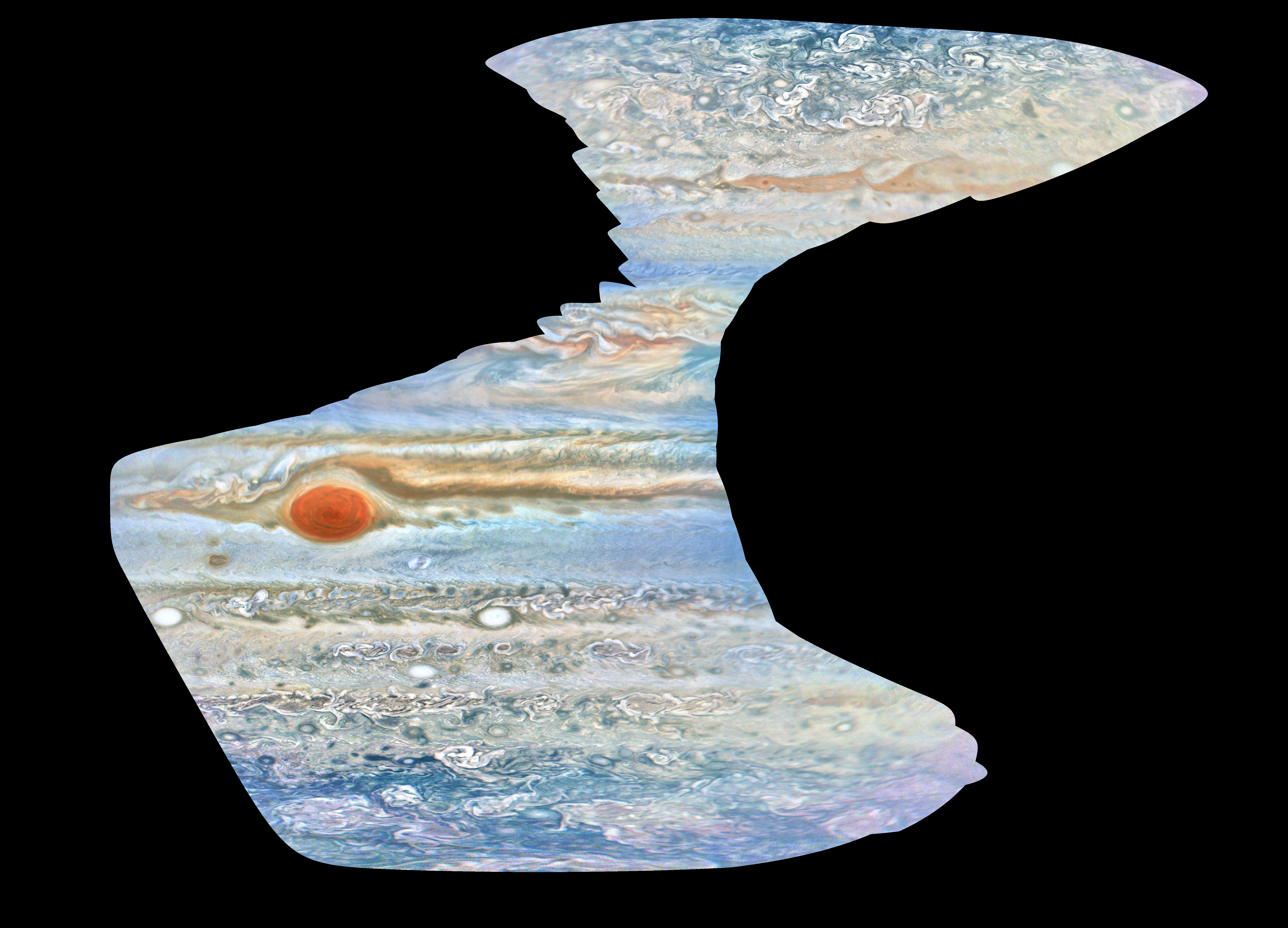}
    \caption{Global mosaic from PJ27 using our processing pipeline. The volunteers are shown square crops from this mosaic.}
    \label{fig:mosaic}
\end{figure}

With the individual map-projected images, we then color correct the images to a uniform scale by applying a histogram equalization, gamma correction ($\gamma = 1.1$) and a high-pass filter to de-trend the global illumination structure. We then stack the images to create a global mosaic, as shown in Figure~\ref{fig:mosaic}. 

\subsubsection{Generating crops for Zooniverse}
\begin{figure}
    \centering
    \includegraphics[width=\columnwidth]{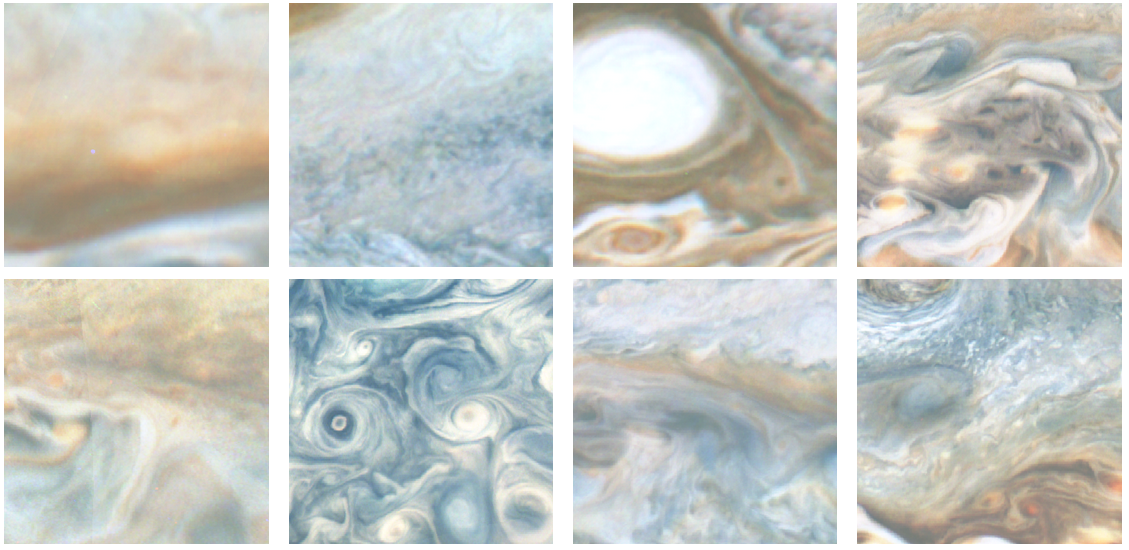}
    \caption{Example of crops generated from our global mosaic. These crops are shown to volunteers on Zooniverse.}
    \label{fig:crop_examples}
\end{figure}

For the Zooniverse project, it is inefficient to show the volunteers the full mosaic, since several vortices form at different length scales, making it inefficient for volunteers to have to comb through large regions for small features. To ensure consistency in what the volunteers observed and also to be consistent with length scales of vortices on Jupiter, we generate equal area crops of the global mosaic. To do so, we randomly sample several coordinate locations on the global mosaic, re-project the region to equal area cartesian coordinates, then crop the region to a size of $7000 \times 7000$ km. We reject all points that are within 1500 km away from a previously sampled point to ensure that the selection doesn't duplicate the same region, but to ensure that large features are sampled several times from different directions. An example of such crops is shown in Figure~\ref{fig:crop_examples}.

\section{Methods} \label{sec:methods}

In the following sections, we describe our citizen science workflow on Zooniverse, and our aggregation and clustering pipeline.

\subsection{Zooniverse workflow}
\begin{figure}
    \centering
    \includegraphics[width=\columnwidth]{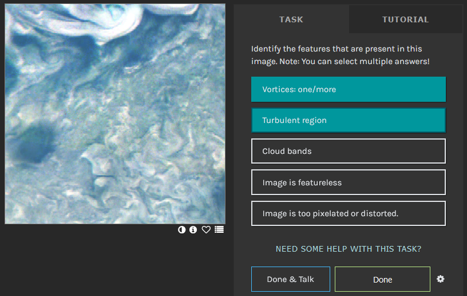}
    \caption{An example of the ``Is there a vortex" workflow, showing the subject on the left and the questions on the right, where we ask volunteers to define the types of atmospheric features in the image.}
    \label{fig:workflow1}
\end{figure}

The Jovian Vortex Hunter Zooniverse project\footnote{\url{https://www.zooniverse.org/projects/ramanakumars/jovian-vortex-hunter/}} workflow was separated into two parts. The first workflow (``Is there a vortex?") is shown in Figure~\ref{fig:workflow1}, where the volunteers are presented with an image segment from \JunoCam{} and asked to describe the type of features seen in the image from a list of:
\begin{itemize}
    \item Vortices
    \item Turbulent region \citep[i.e., Folded Filamentary Region or FFR,][]{Orton2017}
    \item Cloud bands
    \item Featureless image
    \item Image is too pixelated or distorted
\end{itemize}

The volunteers are asked to select one or more of these options. Each image is seen by a maximum of 10 volunteers before being retired from the subject pool, where a subject is an image on the Zooniverse platform that needs to be seen by the volunteer. Any image which has been seen at least 7 times by volunteers and where at least 70\% of the volunteers agree on the ``Vortices" label, is moved to the second workflow for annotation of the vortex ellipses (i.e., images with the `Vortices' label are retired with a minimum consensus of 70\% after at least 7 votes, compared to other labels which are retired after a maximum of 10 votes).

\begin{figure}
    \centering
    \includegraphics[width=\columnwidth]{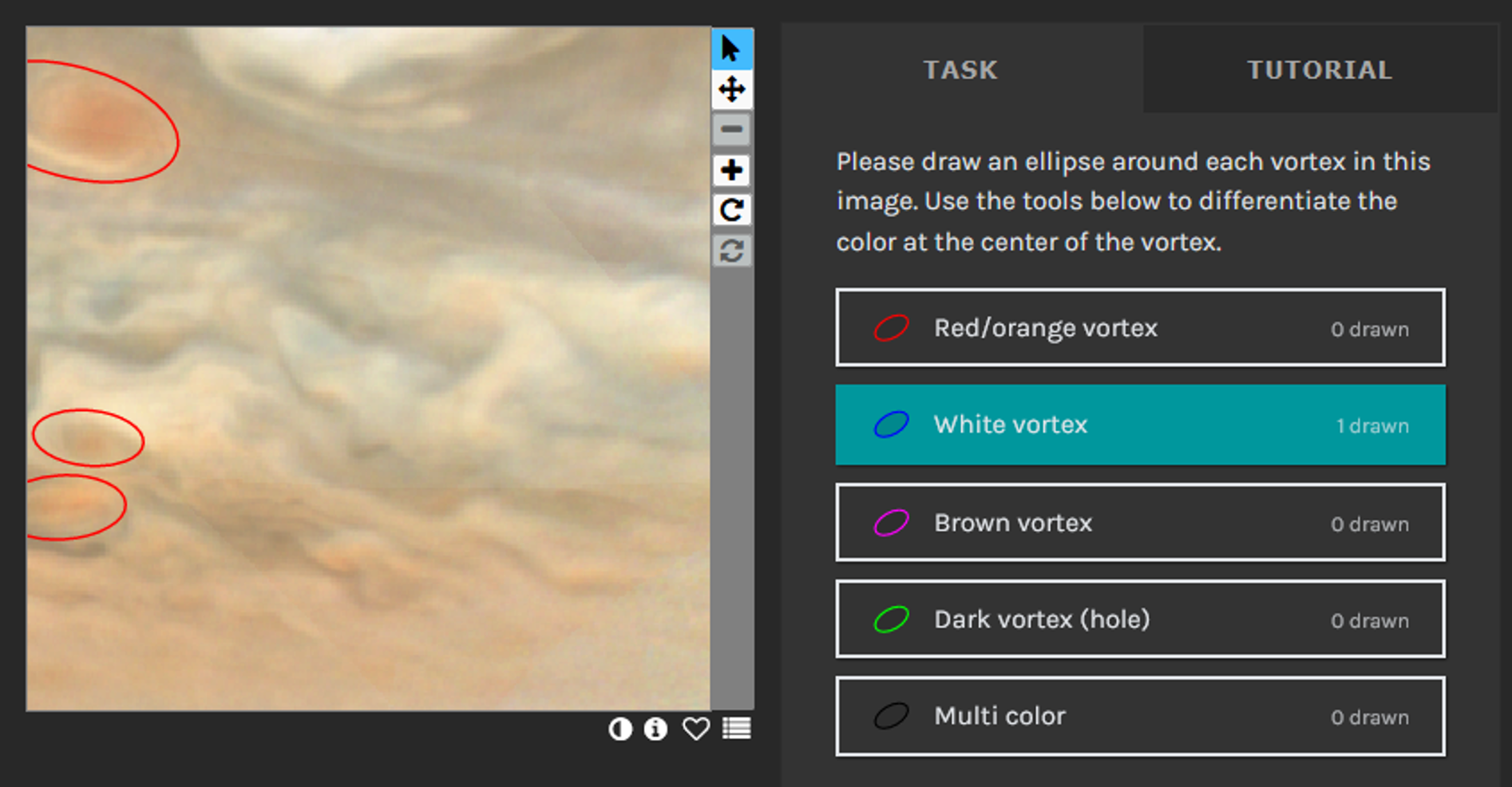}
    \caption{An example of the "Circle the Vortex" workflow, showing the subject on the left and the ellipse annotation tools on the right, where we ask volunteers to annotate vortices based on their color.}
    \label{fig:workflow2}
\end{figure}

The second workflow (``Circle the Vortex") is shown in Figure~\ref{fig:workflow2}. In this workflow, volunteers are shown an image with at least one vortex and asked to annotate it based on its color. The volunteers are provided with different ellipses with the option of white, red, brown, dark and multi-color, and asked to choose the color that best matches the center of the vortex. In the case of multi-color vortices, the volunteers are given a secondary task to label the interior and exterior colors of the vortex (from a choice of white, red or brown). Since this is a much more challenging task compared to the first workflow, each image is seen by 12 volunteers before being retired from the subject pool. Note that the volunteers were free to work on either workflow at any point.

The project was launched in 20 June 2022 and was completed in 3 December 2023, with more than a million classifications by more than 5000 registered volunteers across the two workflows. Zooniverse features a forum-like system called Talk, where volunteers can provide additional information about the image that they classified (e.g., any interesting or unusual artifacts, or questions about the classification), and the ability to tag the image with various qualifiers (e.g.,  \#vortex, \#interesting, \#artifact, \#great-red-spot). Over the duration of the project, we received over 18000 comments on Talk. Of particular interest is the unique tags that were generated in the project, such as `\#compact-red-nursery' which was given to small red vortices forming inside filamentary structures or brown barges. Due to the high resolution and long cadence afforded by the JunoCam images, this presented a valuable opportunity to characterize these types of vortices in good detail. A further study on these unique vortices is planned. 

\subsection{Aggregation}
Given that we use data from several volunteers to draw consensus on a given image, we need to aggregate the multi-volunteer data to a single result per image. This process is different for the first and second workflow due to the task variety between the workflows. We detail our aggregation pipeline below. 

\subsubsection{Workflow 1: Is there a vortex?}
For the first workflow, the volunteers' responses were reduced to a consensus score for each class, which for label $c$ is defined as,
\begin{equation} \label{eq:consensus}
    {\rm consensus}_c = \dfrac{N_c}{N},
\end{equation}
where $N$ is the total number of classifications (i.e., an individual annotation by a volunteer) of that image and $N_c$ is the number of volunteers who selected the label $c$. A consensus of $1$, therefore, means that every volunteer who saw the image labeled it as $c$, while a score of $0$ is when no volunteer chose $c$. 

\subsubsection{Workflow 2: Vortex annotation clustering}
For the second workflow, volunteer annotations are aggregated by clustering the ellipses drawn into separate vortices per subject image. Each ellipse is defined by five parameters: the center ($x$, $y$ in pixel coordinates), radii ($r_x, r_y$, also in pixel coordinates) and angle from the horizontal axis (in pixel space). The clustering is done using the DBSCAN algorithm \citep{Ester1996}, which identifies clusters of points based on their mutual reachability distance defined by a metric. For our analysis, we use the Jaccard metric to find the `distance' (i.e., overlap) between two ellipses drawn by volunteers.
For two ellipses $A$ and $B$, the Jaccard distance is defined as the ratio of the total area of intersection over the total union area (i.e., the intersection-over-union, IoU), as follows,
\begin{equation}
    {\rm Jaccard} = 1 - \dfrac{A \cap B}{A \cup B}.
\end{equation}

Therefore, any two ellipses with perfect overlap will have a Jaccard distance of 0, while those without any overlap will have a Jaccard distance of 1. The DBSCAN algorithm determines the Jaccard distance between all pairs of ellipses in the image and clusters ellipses together which form part of the same `neighbourhood'. A `neighbourhood' is defined using a parameter $\epsilon$, which defines the minimum distance below which ellipses must be automatically considered to be part of the same cluster. In our case, we set $\epsilon = 0.5$, i.e. there must be a 50\% overlap in the area between any two volunteer drawn ellipses for them to be considered part of the same cluster, and thus all pairs of ellipses which share a mutual minimum of $50\%$ overlap with each other will form a single cluster. We also require that any cluster be made up of at least four volunteer drawn ellipses. We note that this is significantly smaller than the retirement limit, since volunteers can annotate multiple vortices in each image, and therefore each ellipse may not contain 12 separate annotations. From our testing, we found that requiring at least four annotations per cluster provided by the best performance between cluster accuracy and completeness of vortices annotated by volunteers.

\begin{figure}
    \centering
    \includegraphics[width=\columnwidth]{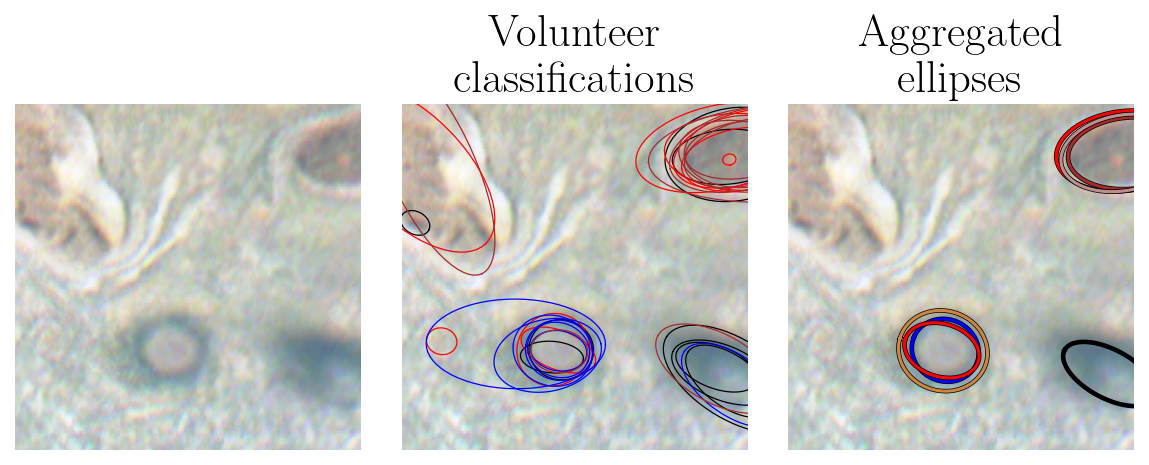}
    \caption{An example of our vortex clustering algorithm. The left panel shows the image shown to volunteers, the middle shows annotations from all volunteers who saw the image, and the right panel shows the resulting clusters. Note that these clusters are done by color, and a further step is to aggregate all the colors for a given vortex. }
    \label{fig:vortex_clustering_wf2}
\end{figure}

Once the clusters are determined, we find an average ellipse per cluster, by minimizing the Jaccard metric between the `average' ellipse and volunteer-drawn ellipses for that cluster, weighted by the probability that the volunteer drawn ellipse is in the cluster. We determine the parameters using the SHGO algorithm \citep{Endres2018}. This process is shown in Figure~\ref{fig:vortex_clustering_wf2}. The ellipses are aggregated separately by color labels. 

For multi-color vortices, we use the same method as above to get the average ellipse, but to determine the color, we calculate the consensus for the volunteer responses for the interior and exterior color of the annotation (which is a sub-task when they use the multi-color annotation tool). The vortex color consensus is calculated in the same way as Equation~\ref{eq:consensus}.

\subsubsection{Matching vortices across Zooniverse subjects}
\begin{figure*}
    \centering
    \includegraphics[width=\textwidth]{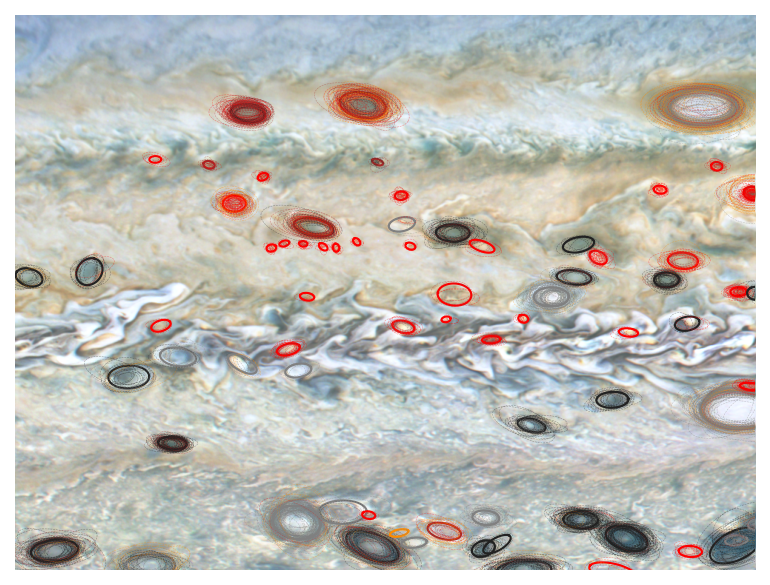}
    \caption{An example from our final clustering step. Each dashed line shows an individual volunteer response across all the images for the same vortex. The final clusters are shown in the solid line with the color corresponding to the color with the most annotations for that vortex.}
    \label{fig:vortex_clustering}
\end{figure*}

The data generation steps assume that there is significant overlap between different images (each image is 7000 km wide, but the separation between images is only 2500 km). Therefore, the same vortex can be seen as many as five times across multiple subjects (each Zooniverse subject is a single JunoCam crop shown to volunteers). To account for this overlap, we project the average ellipses on a longitude-latitude coordinate system and check for overlapping ellipses using the Jaccard metric. If ellipses overlap more than 10\% of their area (i.e., have an IoU greater than 0.1), they are merged and the average ellipse is re-determined for the new cluster of ellipses. This is shown in Figure~\ref{fig:vortex_clustering}. We further used JUPOS positioning data\footnote{\url{http://jupos.privat.t-online.de/index.htm}} and found that very few vortices (outside the GRS, Oval BA and the chain of White Ovals near 40$\degree$ S) were noticeably repeated across multiple perijoves, and thus did not significantly affect the statistical interpretations of the derived vortex properties.
JUPOS is a citizen science initiative to observe Jupiter and track features across multiple amateur observations and determine their drift rates. 

\section{Results} \label{sec:results}

\subsection{Volunteer consensus and accuracy}
\begin{figure}
    \centering
    \includegraphics[width=\columnwidth]{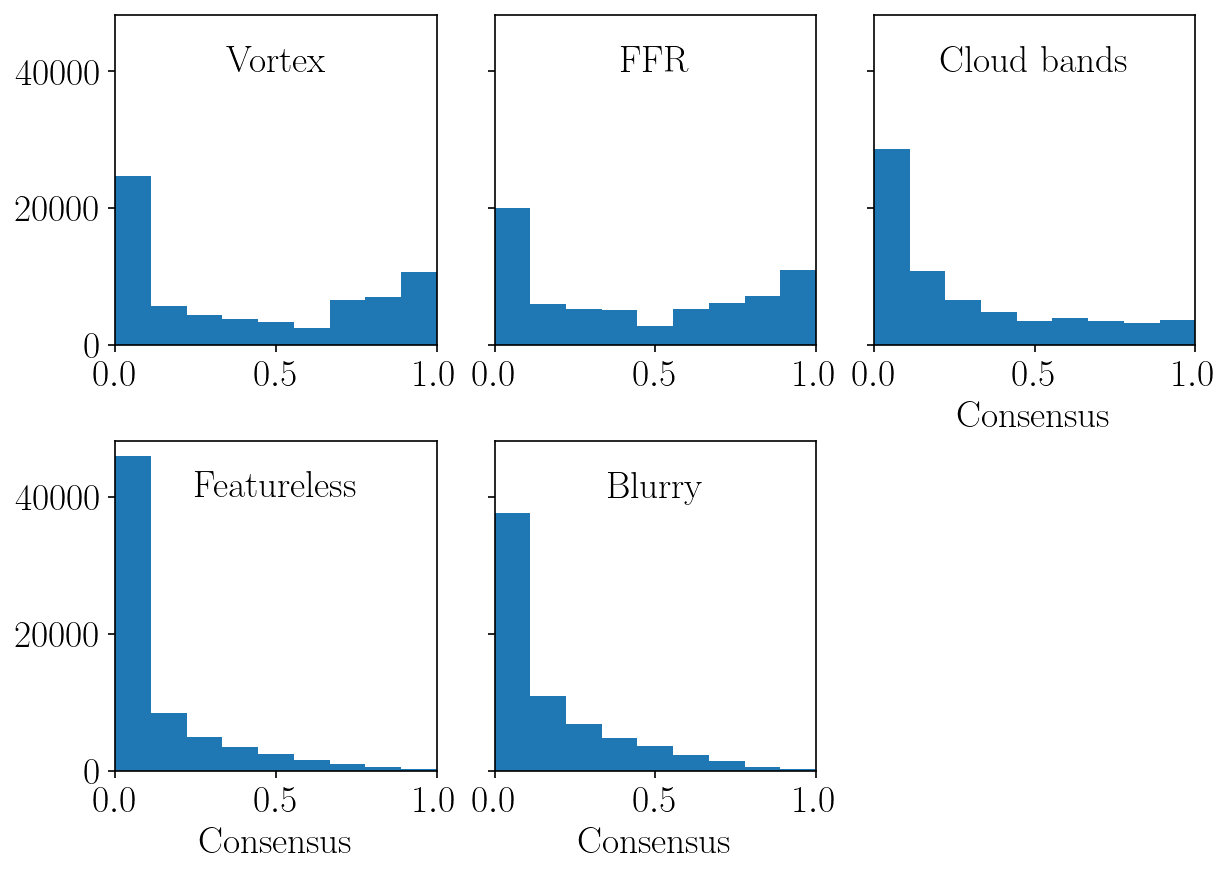}
    \caption{Volunteer consensus for the different atmospheric features in the image. A value of 0 means that no one chose that feature for the image, while a value of 1 means that every volunteer who saw the image labelled it as that feature.}
    \label{fig:consensus_wf1}
\end{figure}

From our project, we find that volunteers are generally very accurate in their labeling of different classes of features in the image, with little training from the research team. A histogram of consensus values for the different labels is shown in Figure~\ref{fig:consensus_wf1}. Vortices and FFRs demonstrate a bi-modal distribution, showing a large fraction of subjects having a clear consensus of a positive label (consensus = 1) or negative label (consensus = 0), while the other three classes show significantly more negative labels with a tail on the consensus distribution. This demonstrates that volunteers can clearly distinguish between the presence and absence of FFRs and vortices in the image.
We attribute these labels to simply be more rare in the dataset (i.e., a small fraction of the images are either featureless or blurry, and given the smaller footprint of JunoCam near the equatorial regions, only a small fraction of the images contain cloud bands). 

\begin{figure}
    \centering
    \includegraphics[width=\columnwidth]{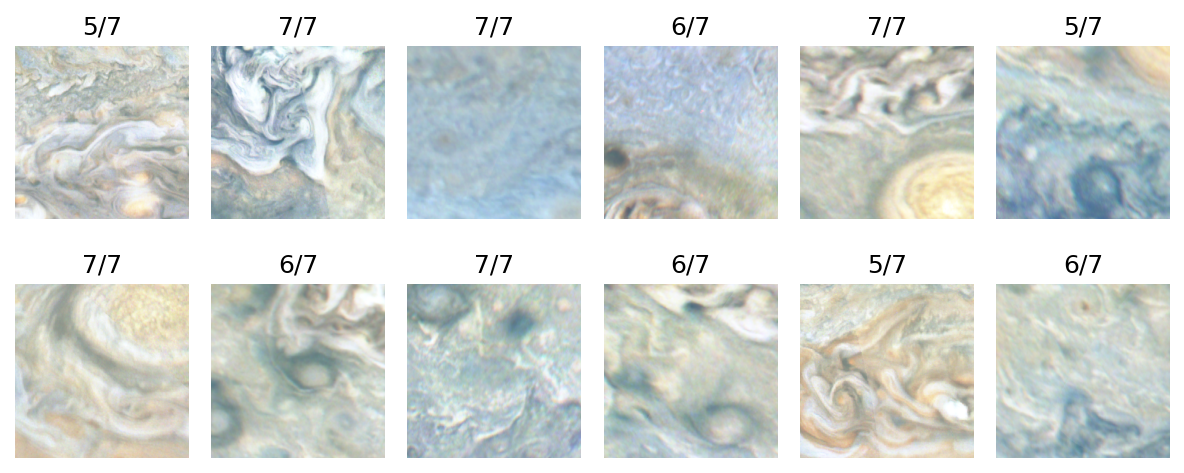}
    \caption{Example of images with high consensus for vortices. All images contain at least one vortex in it. The number above each image shows the fraction of volunteers who classified the image as a vortex over the number of volunteers who saw the image.}
    \label{fig:best_vortices}
\end{figure}

\begin{figure}
    \centering
    \includegraphics[width=\columnwidth]{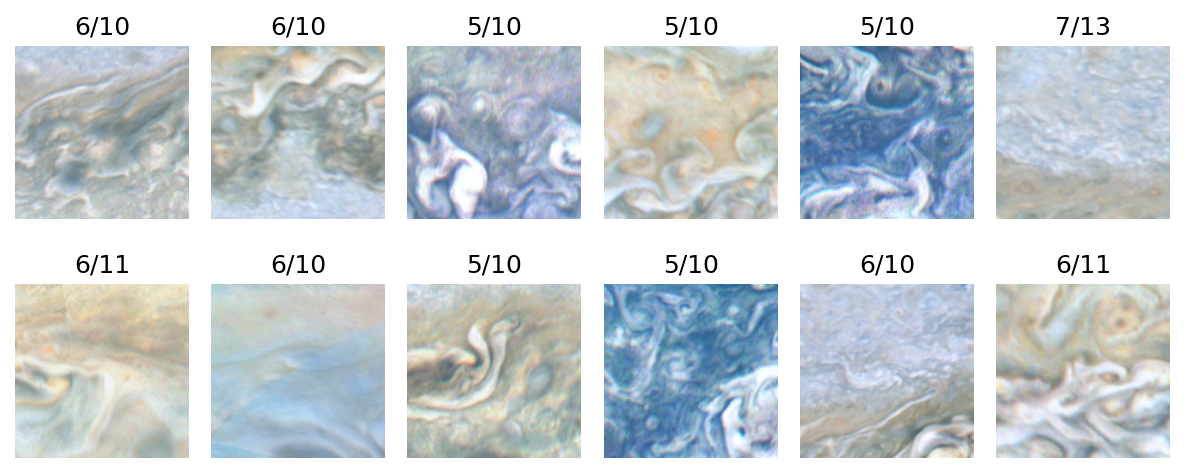}
    \caption{Examples of images with poor consensus for vortices. Most of these examples do not contain vortices, but sometimes there is confusion from the volunteers between FFRs and vortices. The number above each image shows the fraction of volunteers who classified the image as a vortex over the number of volunteers who saw the image.}
    \label{fig:confusing_vortices}
\end{figure}

We found that volunteers very easily identified vortices with well defined edges that showed marked color gradients from the interior to the exterior. An example of images with high agreement between volunteers for vortices is shown in Figure~\ref{fig:best_vortices}. Volunteers were quick to identify vortical features even at very small scales (e.g., the small red vortex in the top row, 3rd image from the left). However, we did find that there was a lot of confusion for vortical features that formed in FFRs. Many volunteers were unsure about whether the swirls in FFRs corresponded to vortices, and there were disagreements between vortices that formed in regions that contained FFRs. Figure~\ref{fig:confusing_vortices} shows images with consensus that was lower than 70\% (i.e. the cut to promote the image to the next workflow). Most of the images do not seem to contain vortices, but there are a few false negatives in these images (e.g., top row, second image from the right and bottom row, image on the far right).

\begin{figure}
    \centering
    \includegraphics[width=\columnwidth]{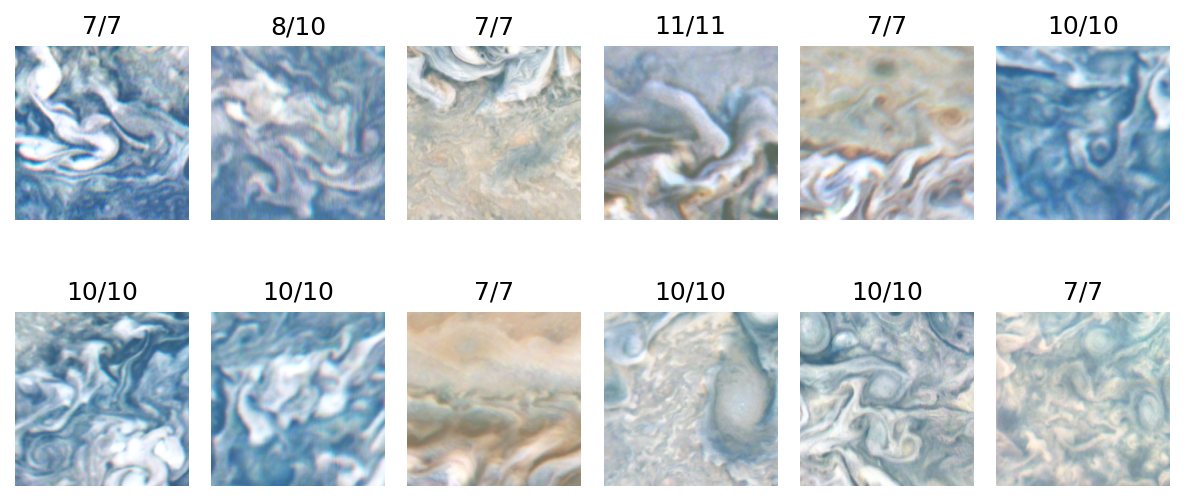}
    \caption{Examples of images with high consensus on FFR features. The number above each image shows the fraction of volunteers who classified the image as containing an FFR over the number of volunteers who saw the image.}
    \label{fig:best_ffrs}
\end{figure}

\begin{figure}
    \centering
    \includegraphics[width=\columnwidth]{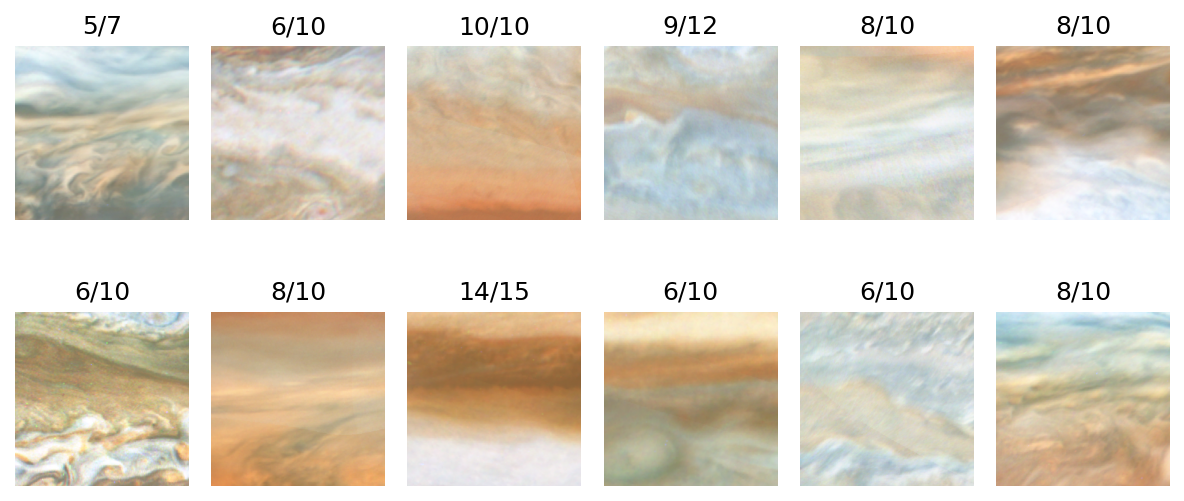}
    \caption{Examples of images with high consensus on cloud bands. The number above each image shows the fraction of volunteers who classified the image as a vortex over the number of volunteers who saw the image.}
    \label{fig:best_cloudbands}
\end{figure}

The volunteers were very good at identifying both the FFRs and the cloud bands. As can be seen in Figure~\ref{fig:best_ffrs}, the high agreement for the ``turbulence'' class all contain marked swirls that are typical of FFRs, and Figure~\ref{fig:best_cloudbands} shows images containing noticeable north-south gradients in color, demonstrating the effect of cloud bands. 

\begin{figure*}
    \centering
    \includegraphics[width=\textwidth]{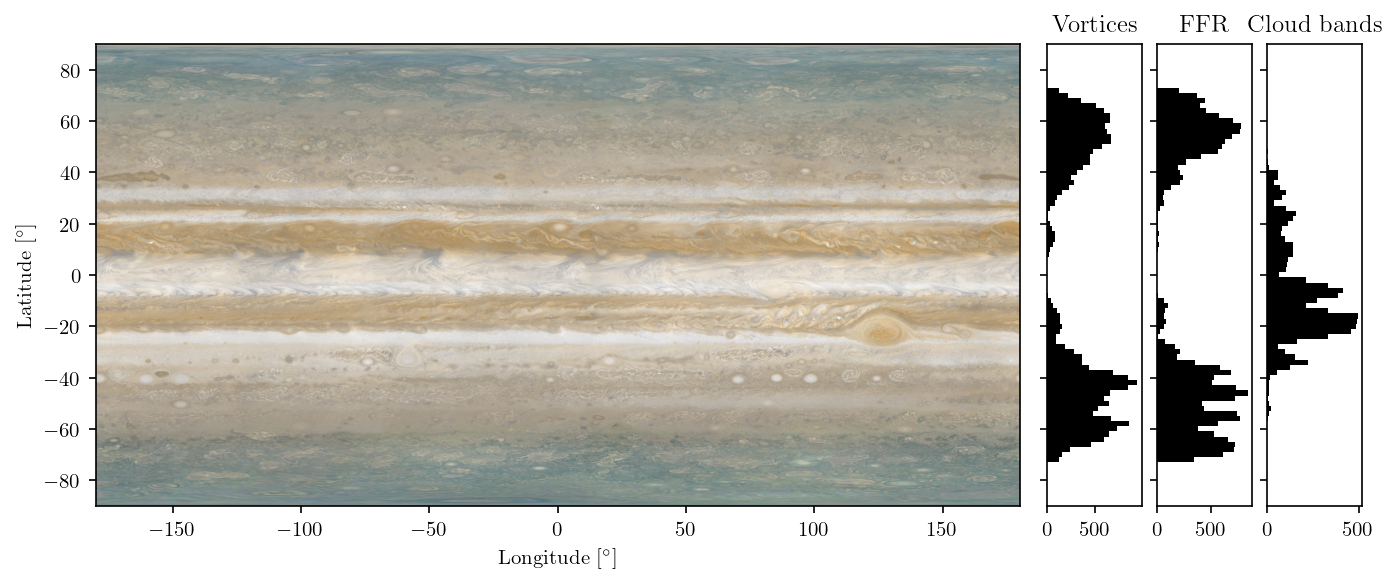}
    \caption{Meridional distribution of vortices, FFRs and cloud bands. As expected, most vortices and FFRs are in the mid-to-high latitudes, while most of the cloud bands are in the lower latitudes. The mosaic on the left is from a composite built using Cassini and \JunoCam{} images (Credit: NASA / JPL-Caltech / SSI / SWRI / MSSI / ASI / INAF / JIRAM / Björn Jónsson)}
    \label{fig:feature_dist_w1}
\end{figure*}

Consequently, we can determine the locations corresponding to each feature (e.g., vortex, cloud band, FFR). This is shown in Figure~\ref{fig:feature_dist_w1}. Note that this distribution does not account for the overlap of the feature over multiple images, and thus the histogram is `smoothed' over a small range of latitudes.

As expected, the vortices and FFRs tend to be dominated in the mid latitudes, while the cloud bands are concentrated closer to the equator. This is a phenomonon defined by the change in the coriolis force with latitude, and correspondingly the Rhines length \citep{Showman2007}, which is the characteristic length scale at which rotational forces dominate over linear advection. At higher latitudes, the Rhines length is larger meaning that smaller vortices remain stable, which results in more vortices forming towards the polar regions compared to the tropics. Therefore, instabilities in the lower latitudes lead to transfer of momentum in the zonal direction, while in the upper latitudes, they lead to the formation of vortices. Images with cloud bands generally peak at the boundary between belts and zones, likely due to it being easier to identify the color gradient and attribute the feature to a band structure. FFRs also seem to be more prevalent near the belt-zone boundaries. 

\subsection{Vortex properties and confidence}

From our second workflow (`Circle the vortex'), we determine the physical attributes of vortices (size, shape, location, color). Volunteers were generally good at annotating the vortices, but there was significant confusion with respect to the color. While this was partly due to there not being a definite color palette provided to the volunteers for matching the vortex colors, this is also due to many vortices not strictly falling in specific color bins. For example, there was significant confusion between dark and brown vortices, since several cyclonic vortices showed a dark patch over the interior that also appeared dark brown in color. The multi-color option was the most difficult to aggregate since many vortices also showed multiple bands of color, making it difficult to accurately pinpoint the interior color, compared to the exterior color. 

However, we find that this disagreement in color is not a downside of the project. Rather, it provides an opportunity to study the spectrum of colors that vortices contain by defining the color agreement score in a fashion similar to the agreement score defined in Eq~\ref{eq:consensus}. For a given aggregated ellipse, we calculate the color agreement for each color as the ratio of the number of volunteer-drawn ellipses of that color, divided by the total number of ellipses for the vortex. In this way, we can separate the vortices with clearly defined color (e.g., white anti-cyclones, the Great Red Spot, etc.) from those with poor color definition (e.g., brown barges with light color, vortices with haze layers, etc.). We define the final color of the vortex as the one with the highest consensus.

\begin{figure}
    \centering
    \includegraphics[width=\columnwidth]{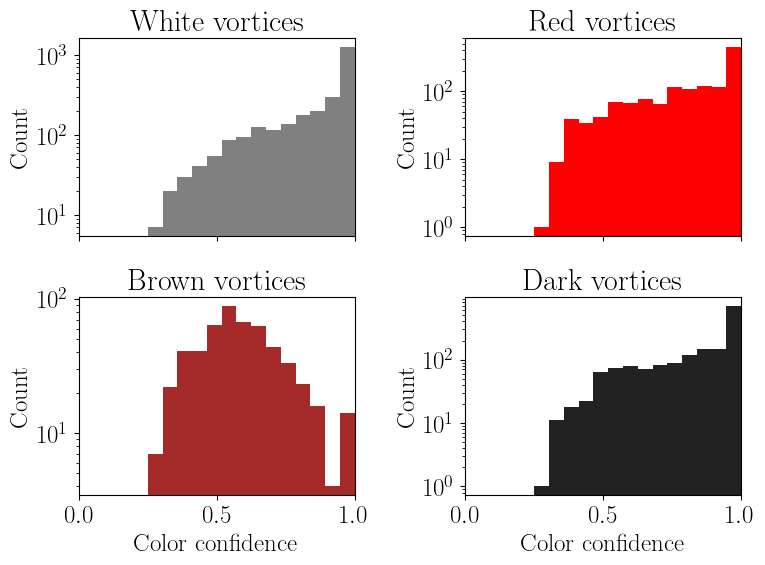}
    \caption{Consensus for the vortex color from volunteer annotations.}
    \label{fig:color_confusion}
\end{figure}

\begin{figure}
    \centering
    \includegraphics[width=\columnwidth]{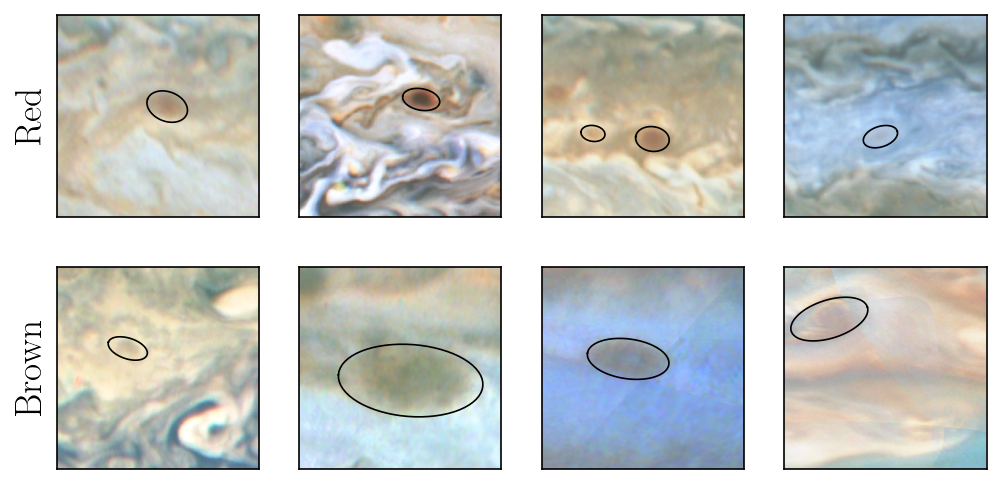}
    \caption{Examples of vortices with confusion between red and brown colors.}
    \label{fig:red_brown_confusion}
\end{figure}

Figure~\ref{fig:color_confusion} shows the distribution of consensus for different colored vortices. White, red and dark vortices are usually clearly identified while volunteers have the most confusion with brown vortices.  Indeed, we found that volunteers had most difficulty discerning between brown and red vortices, since several features had either a dark-red, or a light-brown color, which fell in between these two color bins. This is shown in Figure~\ref{fig:red_brown_confusion}, which provides examples of low confidence brown and red vortices, where we see the morphological similarities in vortices of these two colors. Particularly, we also find that brown vortices on Jupiter are not limited to barges, but can also form very small vortices that contain a deep brown hue. Given a lack of a fixed color palette at the start of this project, we find that care must be taken for brown vortices with low color consensus.

\begin{figure}
    \centering
    \includegraphics[width=\columnwidth]{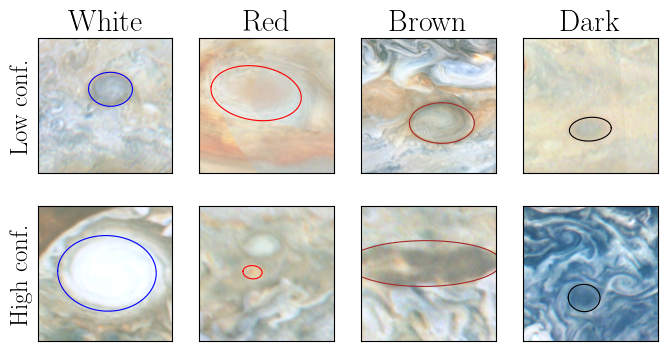}
    \caption{Examples of vortices with low (top row) and high (bottom row) color confidence for each color class.}
    \label{fig:color_confusion_examples}
\end{figure}

Figure~\ref{fig:color_confusion_examples} shows typical examples of vortices with low and high color confidence for each color of vortices. Volunteers typically faced difficulties when vortices had multiple color characteristics (e.g., when the core of the vortex was a different color from the edges, or when the vortex color itself was weak and was affected by the color of the background clouds).

\begin{figure}
    \centering
    \includegraphics[width=\columnwidth]{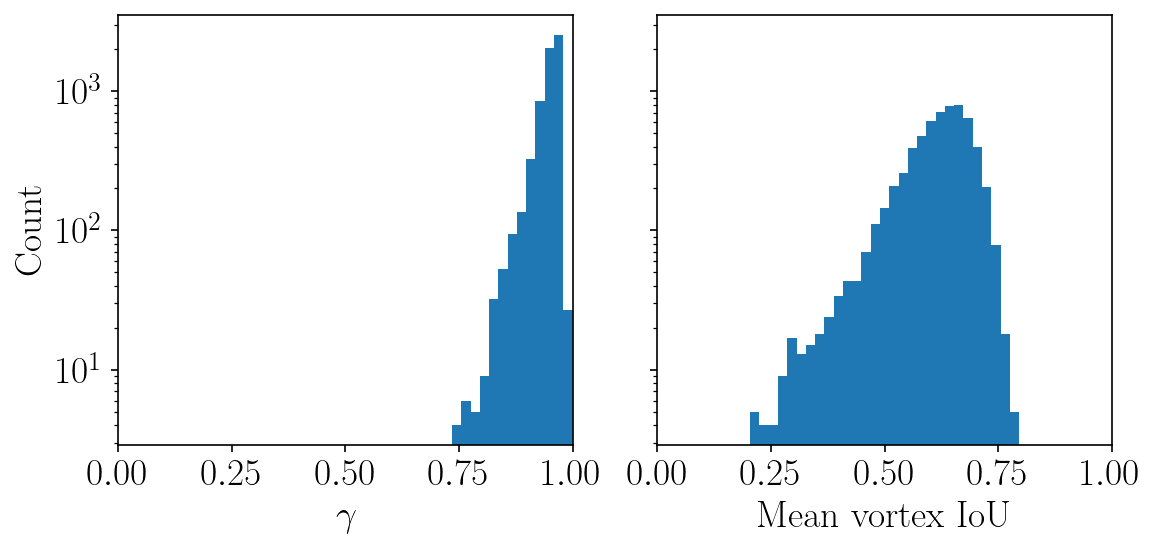}
    \caption{Vortex-ellipse-clustering confidence. The left panel shows the 1-$\sigma$ error in the ellipse scale (see text for details), where values close to 1 correspond to low error and values close to 0 correspond to high error. The right panel shows the average IoU of the clustered ellipse with respect to individual volunteer annotations. }
    \label{fig:shape_confidence}
\end{figure}

The shape parameters are generally better defined. We find that the biggest confusion in the vortex morphology comes from distinguishing between swirls in FFRs and vortices, with many volunteers circling filaments as vortices. However, given the high retirement rate (number of volunteers needed to aggregate consensus on an image), it is possible to filter these out by setting a high classification threshold on a given vortex. We also find that there is no discernable difference between the confidence and/or IoU of the average ellipse for different colors; volunteers are easily able to determine and agree on the shape of the vortex. Figure~\ref{fig:shape_confidence} shows the 1-$\sigma$ error in the ellipse scale and the distribution of mean IoU of the average ellipse with respect to volunteer classifications. $\gamma$ defines the lower and upper bound scale of the ellipse to a 1$\sigma$ confidence (i.e., a $\gamma$ of 0.5 means that the lower bound on the ellipse is half the scale of the average and the upper bound is twice the scale of the average). A value close to 1 signifies a low $\sigma$. The IoU distribution peaks at a value of around 0.6 (i.e., on average each vortex has an overlap of 60\% with individual volunteer annotations). 

\begin{figure}
    \centering
    \includegraphics[width=\columnwidth]{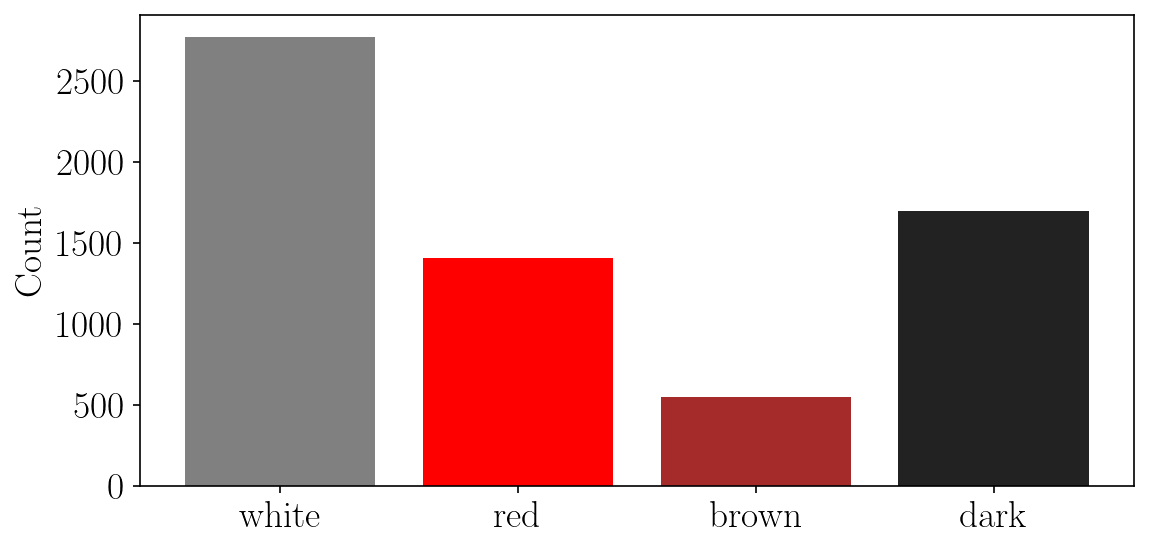}
    \caption{Number distribution for vortices of different colors from our study.}
    \label{fig:color_distribution}
\end{figure}

At the end of our filtration stages, we have 7222 vortices, where 3063 were labeled white, 1596 were red, 639 were brown and 1881 were dark (Figure~\ref{fig:color_distribution}). Only 43 vortices had a multi-color label. For this paper, we will focus mainly on the former four categories, since it is difficult to draw large conclusions from the small population of multi-color vortices. 

\subsection{Distribution of vortex properties}
\begin{figure}
    \centering
    \includegraphics[width=\columnwidth]{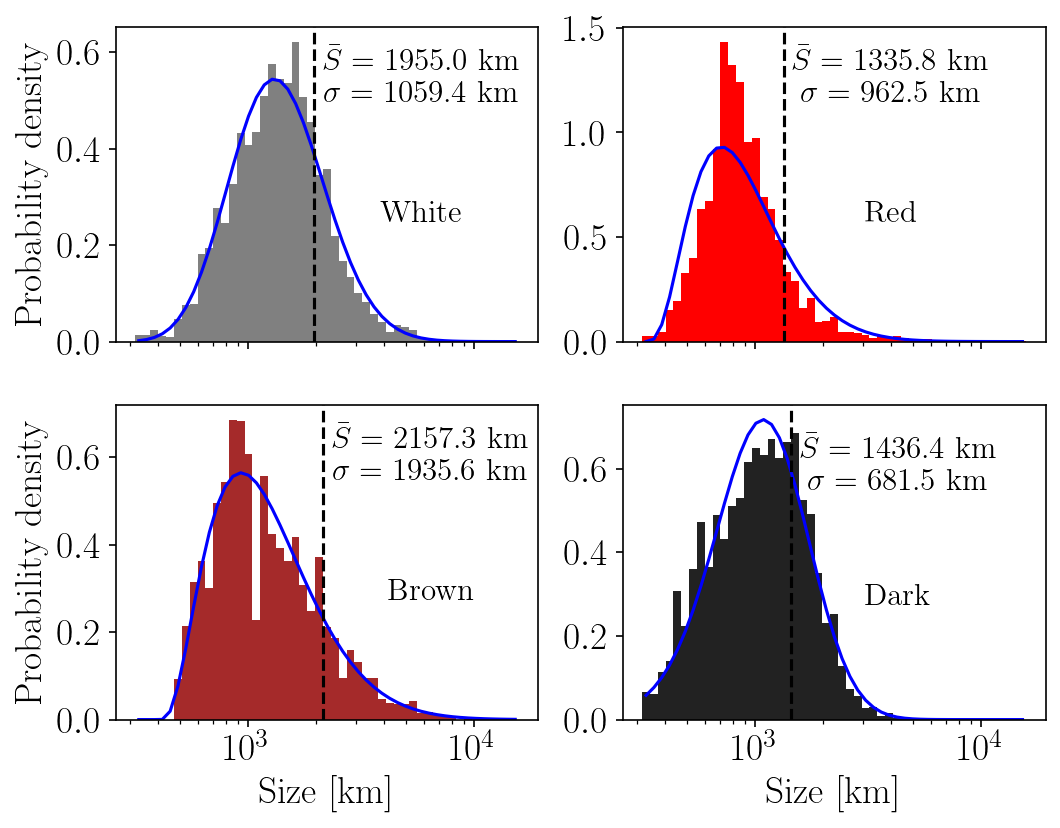}
    \caption{Distribution of vortex sizes by color. The blue line shows the log-normal fit to each distribution with the mean and standard deviations annotated. Note that the y-axes have different scales for each color.}
    \label{fig:size_dist}
\end{figure}

Figure~\ref{fig:size_dist} shows the distribution of vortex size (semi-major axis) for different vortex colors, along with the mean value. The blue line in each shows a fit with a log-normal distribution. In general, the vortex size is well represented by a log-normal distribution, which is consistent with studies of eddies in Earth's oceans \citep{Tang2019}. The white and dark vortices are much better fits to these distributions, compared to red and brown vortices. In particular, the red vortices show the poorest fit to the distribution, possibly due to being made up of multiple different populations of vortices (e.g., groups of large red cyclones and groups of small vortices which form within FFRs).
Across the colors, the mean vortex size is between $1500-2000$ km, and most of the vortices are between $300$-$10000$ km in size. The white and brown vortices occupy the larger end of the distribution while the red and dark vortices are much smaller.

\begin{figure}
    \centering
    \includegraphics[width=\columnwidth]{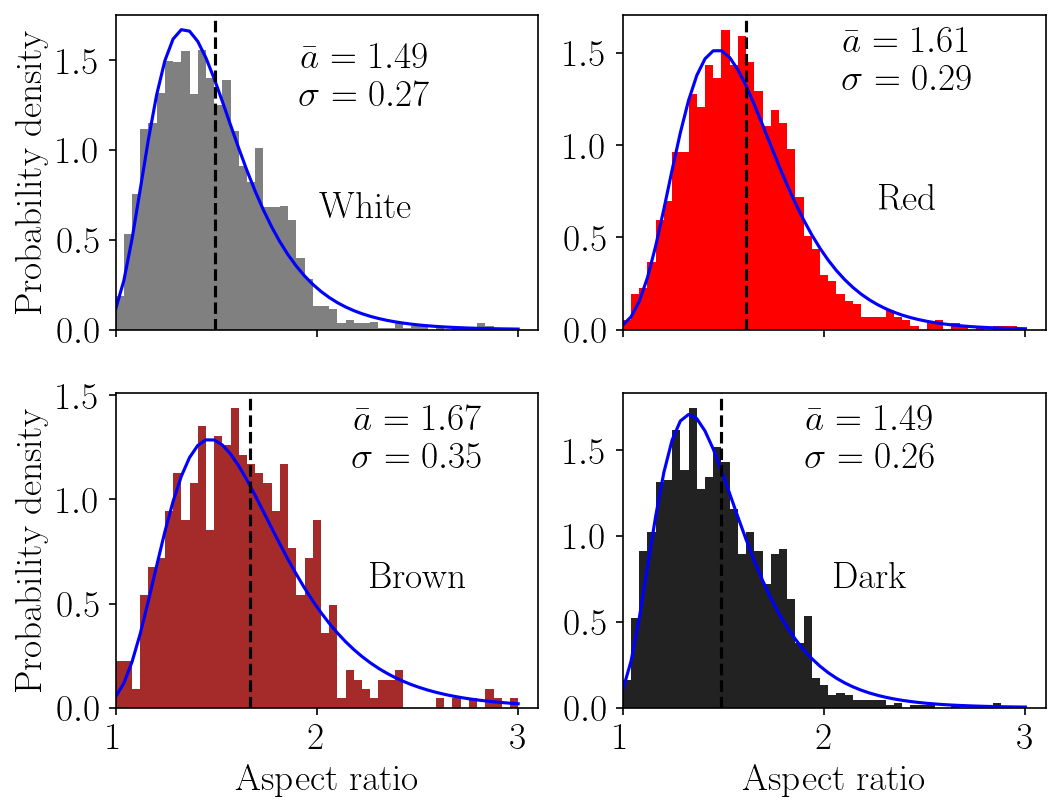}
    \caption{Distribution of vortex aspect ratios by color. The blue line shows the log-normal fit to each distribution with the mean and standard deviations annotated. Note that the y-axes have different scales for each color.}
    \label{fig:aspect_ratio}
\end{figure}

Figure~\ref{fig:aspect_ratio} shows the distribution of vortex aspect ratio (ratio of semi-major to semi-minor axis in physical distance units) for different vortex colors, along with the mean value. Compared to the size distribution, the aspect ratio for all four vortex colors is much better approximated by the log-normal distribution, with a mean of around $1.5$, except for brown vortices, which have a mean of $1.6$ and a much shallower tail towards higher aspect ratios. This is due to the existence of a large fraction of brown barges, which are generally extended features. 

\begin{figure}
    \centering
    \includegraphics[width=\columnwidth]{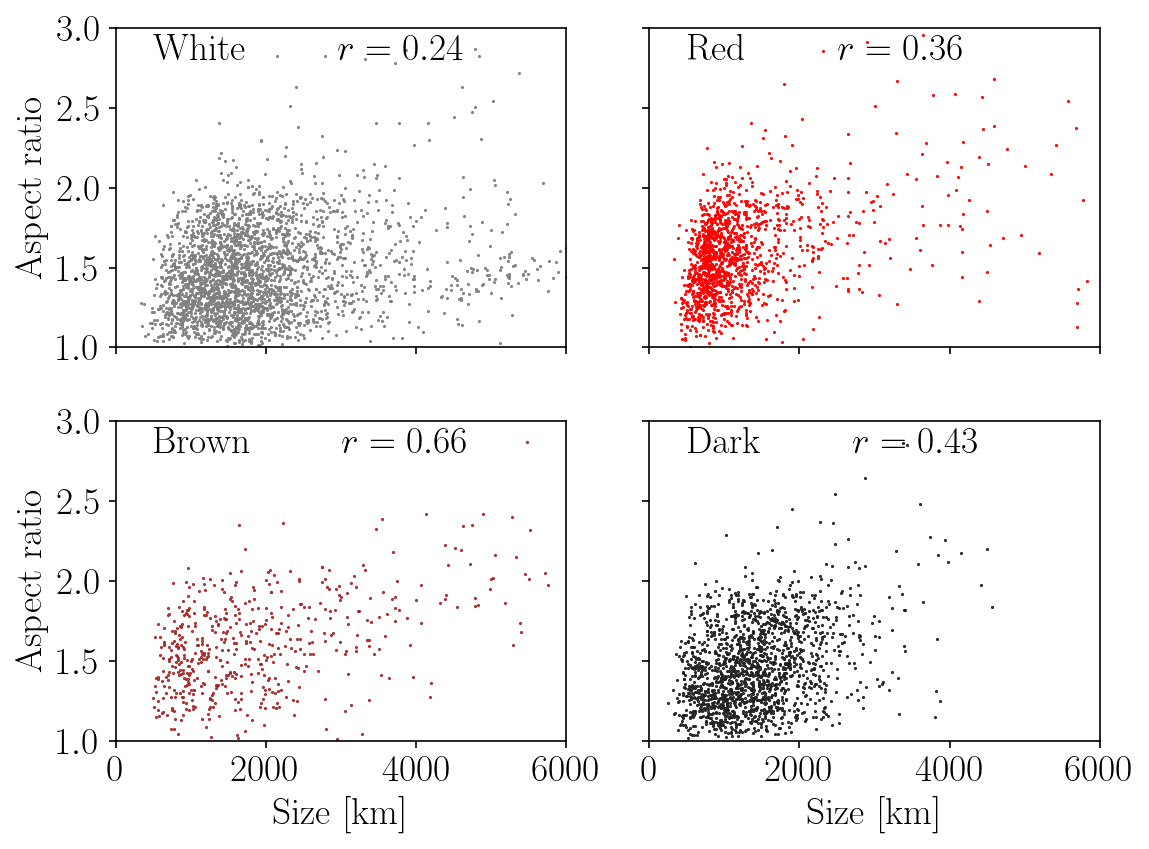}
    \caption{Distribution between vortex size and aspect ratio. There is no clear trend observed with any of the colors, showing that sizes and aspect ratios are not correlated quantities.}
    \label{fig:ratio_size_corr}
\end{figure}

Figure~\ref{fig:ratio_size_corr} shows the relation between vortex size and aspect ratio for the different colors. Most of the vortices do not show a strong correlation (i.e., Pearson's correlation coefficient, $r < 0.5$). In fact, the white and red vortices show almost no trend in aspect ratio with size, and only a few large vortices appear to be strongly elliptical. The exception to this are brown vortices, which show a non-negligible trend ($r = 0.66$).

\subsection{Meridional distribution of vortices} \label{sec:lat_dist}
\begin{figure}
    \centering
    \includegraphics[width=\columnwidth]{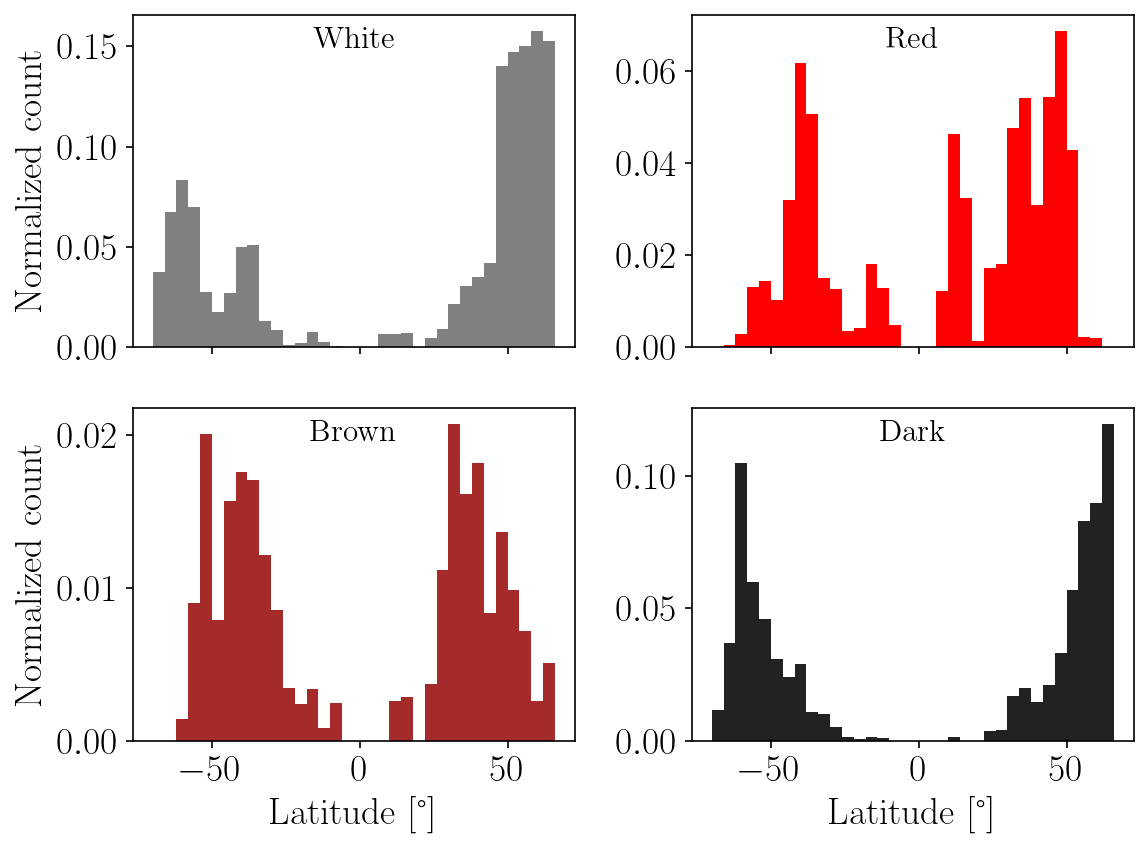}
    \caption{Distribution of vortices by latitude, normalized by the total number of images at each latitude bin. }
    \label{fig:vortex_lat_dist}
\end{figure}

Figure~\ref{fig:vortex_lat_dist} shows the distribution of vortices as a function of latitude by color. Since the \JunoCam{} footprint is variable with latitude, we normalize the raw count per latitude bin by the number of \JunoCam{} images that the volunteers saw in that bin. Therefore, the histogram describes the fraction of vortices of that color for a given latitude bin assuming a constant distribution with longitude. It is evident that the distribution of vortices is vastly different with color: dark and white vortices are strongly concentrated towards the higher latitudes. The small peak at $-40\degree$ for white vortices corresponds to a collection of White Ovals (WO) which are long-lived white anti-cyclones at that latitude.  

\begin{figure}
    \centering
    \includegraphics[width=\columnwidth]{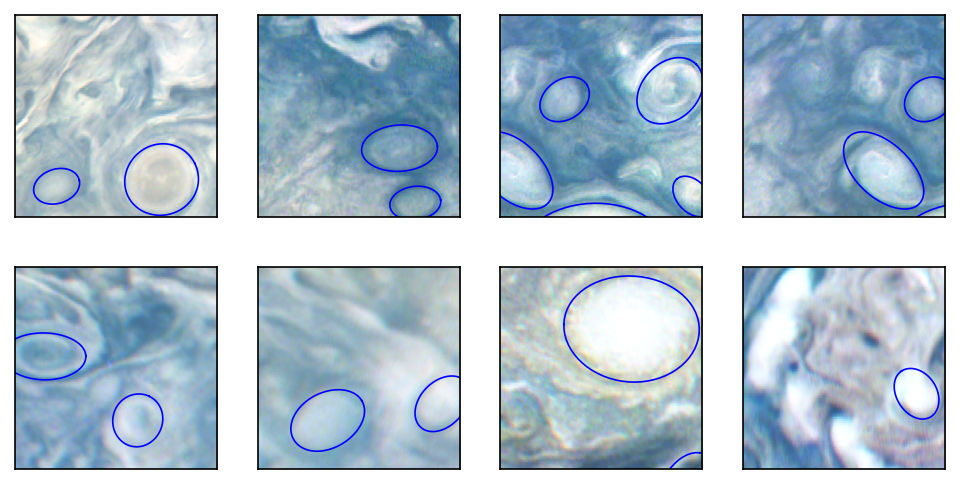}
    \caption{A selection of white vortices in the polar region (the top panel denotes northern hemispheric ovals, while the bottom panel is for southern hemisphere).}
    \label{fig:white_polar_ovals}
\end{figure}

It is also interesting to see the asymmetry between the north and south polar regions for the white ovals. Since this is normalized by the number of available images in each latitude bin, we are confident that this is not an observer bias, but rather is a real trend in the data. We find that most of the white ovals found by volunteers in these polar regions are either vortical structures in filamentary arms or bright spots in FFRs (Figure~\ref{fig:white_polar_ovals}). As seen in Figure~\ref{fig:feature_dist_w1}, there is a remarkably more homogeneous distribution of FFRs in the northern hemisphere, compared to the southern hemisphere, where FFRs tend to form along specific latitudinal bands. However, since these images are a snapshot in time, we do not know whether a fraction of these volunteer detected vortices in the north are truly stable long-term, or whether they will morph into non-vortical structures in a short timescale. Nonetheless, even given the sample shown in Figure~\ref{fig:white_polar_ovals}, we do see that most (if not all of them) appear to have vortical characteristics to the eye, but dynamical information is needed to disentangle FFRs from vortices.

Interestingly, brown and red vortices preferentially form in the mid latitudes. Brown vortices tend to be more common in the mid latitudes, between $30$ and $50\degree$ north and south latitudes, which is consistent with prior detection of brown barges \citep{MoralesJuberias2002}. Red vortices, on the other hand, are not homogeneously distributed, and instead form at specific latitudinal bands. Therefore, each vortex is coupled strongly to the local dynamical conditions of the atmosphere, and possibly to local chemical abundance. 

\begin{figure}
    \centering
    \includegraphics[width=\columnwidth]{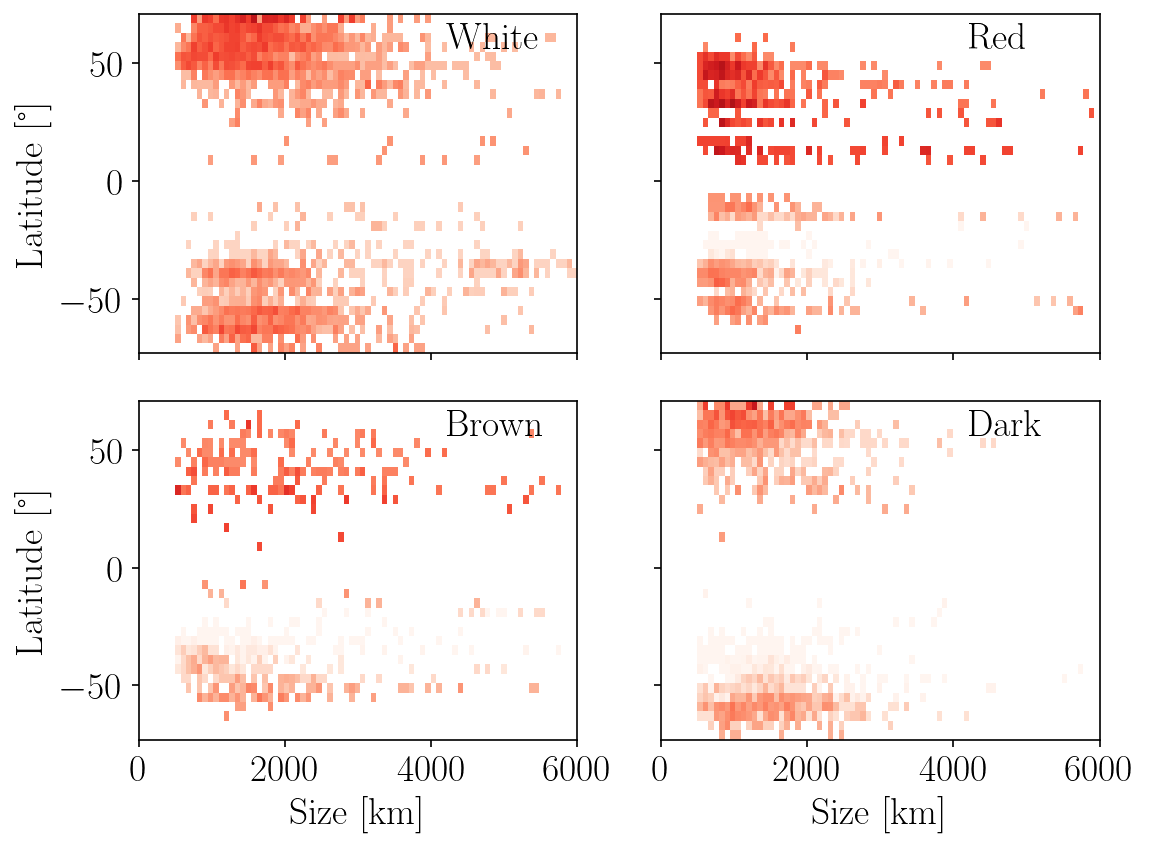}
    \caption{Distribution of vortex sizes by latitude. The histogram is weighted by the inverse of the number of images in each latitudinal bin in order to remove observational bias.}
    \label{fig:size_lat}
\end{figure}

Figure~\ref{fig:size_lat} shows the distribution of vortex sizes as a function of latitude with a running mean showing the average trend, as a function of color. In general all vortices show a trend going from small sizes near the poles to larger sizes near the equator, although this is most prominent for the white and dark vortices. 

\subsection{Relation of vortex properties to atmospheric dynamics}

Vortex evolution on Jupiter is complicated and governed by several independent processes including multi-scale convection (dry and moist) and fluid dynamical instabilities \citep{Brueshaber2019,Inurrigarro2022}. To investigate the formation mechanism of these vortices, we study the relation between the vortex properties and corresponding state of the atmosphere at the location that they are observed. Note, however, that it is difficult to directly relate the distribution of the observed state of the vortices to both the underlying atmospheric structure and/or its evolutionary track, since, statistically, these vortices can be observed at any point during their evolution. As shown in Section~\ref{sec:lat_dist}, different classes of vortices have different distributions with latitude, and therefore, likely follow different evolutionary tracks. 

Vortex formation is inherently tied to specific instability mechanisms in the atmosphere, which have several sources. Primarily, the cause of dynamical instabilities is from the unstable growth of small scale (horizontal and/or vertical) perturbations. Using perturbation theory, we can derive conditions for the state of the atmosphere and the nature of the perturbation which would lead to unstable growth. Such criteria include Rayleigh's inflection point theorem \citep{Rayleigh1880} which states the fluid is stable if the vorticity gradient does not change sign, and the Charney-Stern criteria \citep{CharneyStern} which states that a fluid is stable if the potential vorticity gradient does not change sign. In this section, we investigate these different stability criteria by determining relations between observed vortex properties and the dynamics of the atmosphere at that location.

\subsubsection{Relation to zonal wind}
Large vortices on Jupiter have shown strong correlation to the local zonal wind profile (e.g., the Great Red Spot sits in a narrow shear region between two alternating jets), and indeed vortices have been shown to be stable if the background vorticity (i.e., meridional gradient of the zonal wind) is similar to the vortex vorticity \citep{Marcus1988}. In this section, we investigate the relation between the location of vortices on the planet derived in this study and the zonal wind profile. 

\begin{figure*}
    \centering
    \includegraphics[width=\textwidth]{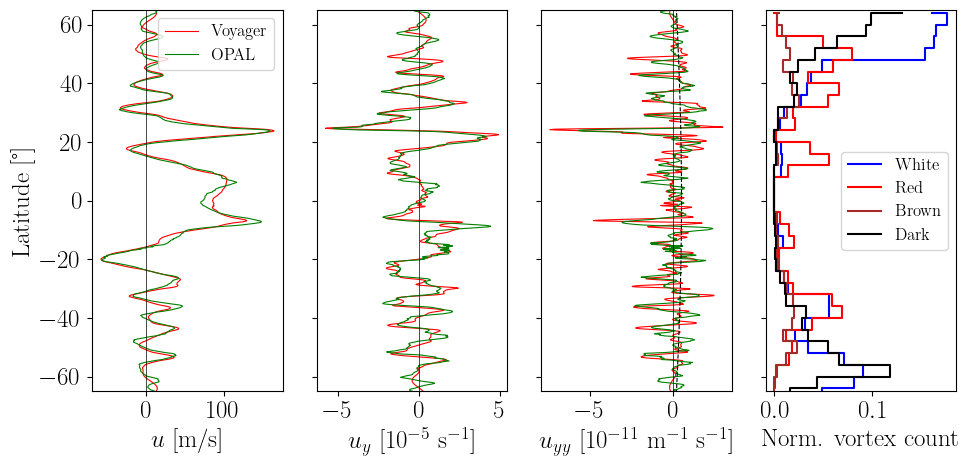}
    \caption{Cloud tracked zonal wind speeds from Voyager (red) and OPAL 2016 (green), and their corresponding vorticity and vorticity gradient. The $u_y$ and $u_{yy}$ profiles are smoothed using a moving average with a window of $0.75\degree$.}
    \label{fig:zonal_wind}
\end{figure*}

We tested different zonal wind speeds derived from previous cloud tracking analyses from the Voyager and OPAL data \citep{Limaye1986,Tollefson2017}. 
Figure~\ref{fig:zonal_wind} shows these two profiles, along with their corresponding vorticity,  vorticity gradient and the distribution of vortices of each color. For both profiles, we applied a moving average on the $u_y$ and $u_{yy}$ profiles (window size of $1\degree$, or 4 points for the Voyager data and 20
points for the OPAL data) in order to remove high frequency noise. 

In general, we find that the distribution of vortices matches somewhat poorly with the zonal wind profile and its corresponding gradients. Specifically, vortices preferentially form towards the poles, where both the background vorticity $u_y$ and vorticity gradient ($u_{yy}$) are smaller, and the peaks in each profile do not necessarily line up with the peaks in the vortex distribution. The exceptions to this are at $\sim 18-20\degree$ N and $\sim 40\degree$ S.

\begin{figure}
    \centering
    \includegraphics[width=\columnwidth]{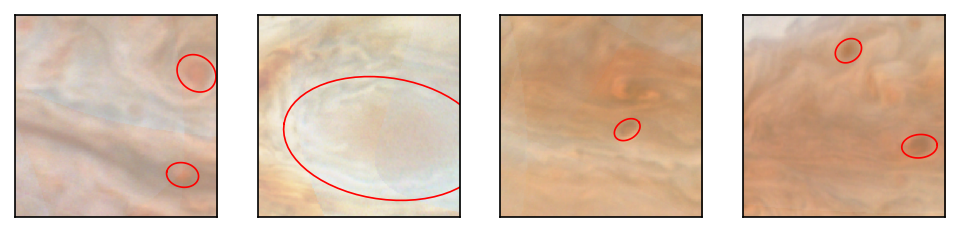}
    \caption{A selection of red vortices between $10\degree$ N and $20\degree$ N.}
    \label{fig:red_vortices_15N}
\end{figure}

In the former case, the increase in red vortices is consistent with the location where $\beta - u_{yy}$ changes signs, given by Rayleigh's inflection theorem \citep{Rayleigh1880}. Figure~\ref{fig:red_vortices_15N} shows a selection of vortices at this location, showing a diversity of different sizes, where most are small cyclones forming inside filamentary structures (panels 1, 3, and 4) with the exception of one large anti-cyclonic oval (panel 2).  

\begin{figure}
    \centering
    \includegraphics[width=\columnwidth]{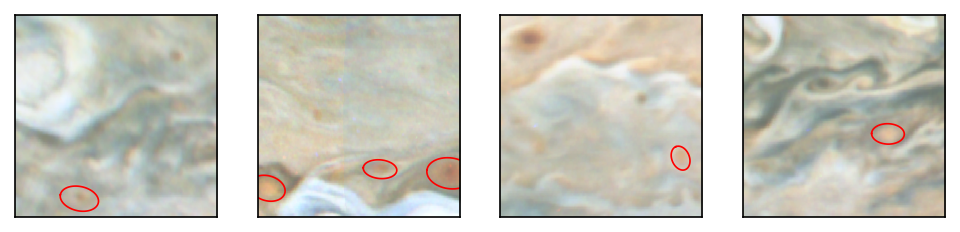}
    \includegraphics[width=\columnwidth]{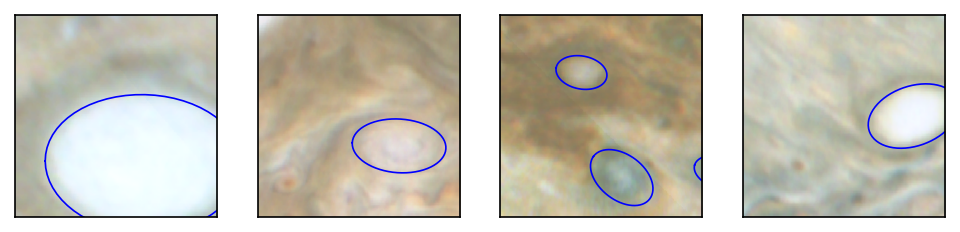}
    \caption{A selection of red (top row) and white (bottom row) vortices between $35\degree$ and $45\degree$ S.}
    \label{fig:vortices_40S}
\end{figure}

In the latter case, there is a peak in both the red and white oval distribution at $40\degree$ S, which is also consistent with Rayleigh's inflection theorem (i.e., peak in the background vorticity). Here, the white ovals form predominantly large vortices, while the red vortices are generally much smaller in sizes. 

Across these samples of vortices, it is generally unclear about the longevity of their structure. Observations have shown the larger ovals are generally stable at these locations \citep[e.g., the vortex at $19\degree$ N has survived since at least 2008, ][]{BarradoIzagirre2021}, but there is very little data on the smaller red vortices. Indeed it seems that the smaller vortices form within filaments, and while they are numerous, they do not seem to persist, since there are very few records of them. 

Rayleigh's inflection theory is simply a necessary condition for instability, and therefore, while it is violated at multiple locations on Jupiter, we retain the long term stability of the zonal jets \citep{Ingersoll1981,Ingersoll2004}. As such, it is interesting that these specific locations are not similarly stable like the rest of the planet, but result in an increase in vortex genesis. It is possible that several of the smaller vortices might be formed convectively due to reductions in vertical static stability within the cylonic eddies in the observed filaments \citep{DowlingGierasch1989,Sankar2021}. For the larger ovals at these locations, modeling studies have shown that they should likely contain deep roots \citep{Palotai2014,Legarreta2008,BarradoIzagirre2021}, which lead to their stability. Since the analysis here is simply 1-dimension (i.e., $u$ as a function of latitude), it might be that we are missing information about the vertical structure of the atmosphere, which lends another dimensionality to the stability of the atmosphere \citep{Dowling1995,Inurrigarro2022}. 

\subsubsection{Atmospheric dynamics and instability mechanisms}
To better understand the distribution of vortices, we must investigate the instability mechanisms in the atmosphere while accounting for the vertical stratification. Despite the fact that instability conditions that are purely based on the zonal wind and vorticity are violated throughout the jovian atmosphere \citep{Ingersoll1981}, despite the presence of several long-lived, stable features on the planet (including several vortices such as the Great Red Spot and Oval BA),it is more likely that the atmosphere is neutrally stable to perturbations through the Arnol'd Second Stability criteria \citep{Dowling1995}, which states that vortex phase velocity must never be larger than the advection velocity. Technically, this means that the inverse of the vorticity Mach number (``Ma''), must satisfy \citep{Arnold1966},
\begin{equation}
    \label{eq:mach_general}
    \dfrac{1}{\rm{``Ma"}} \approx -L_d^2 \dfrac{dq}{d\Phi} < 1,
\end{equation}
where $L_d$ is the Rossby deformation radius, which is related to the vertical static stability given by the Brunt-V\"ais\"al\"a frequency, $\Phi$ is the streamfunction and $q$ is the potential vorticity. We can expand Eq~\ref{eq:mach_general} \citep{Dowling1995} giving,
\begin{equation}
    L_d^2 \dfrac{\beta - u_{yy} + (1/L_d^2) (u - u_d)}{u - \alpha} < 1,
\end{equation}
where $u$ is the zonal wind speed, $u_{yy}$ is the meridional relative vorticity gradient, $u_d$ is the zonal wind at depth and $\alpha$ is used to perform a Galilean shift in the frame of reference to satisfy the inequality. Voyager data showed that $\alpha \sim 0$ \citep{Dowling1995} and over the vertical scale of the observed vortices, vertical shear is negligible, and therefore $u_d \sim u$. Therefore, to be neutrally stable, we require, that the critical value of $L_d$,
\begin{equation}
    \label{eq:ldc}
    L_{d, c}^2  \lesssim \dfrac{u}{\beta - u_{yy}}.
\end{equation}

Since this is a necessary condition for fluid stability in the atmosphere, for $L_d < L_{d, c}$, the atmosphere will be stable to perturbations, but we require $L_d > L_{d, c}$ for instabilities to occur.  Note that since this calculation is based on the quasi-geostrophic assumption, we are excluding effects that promote or impede stability within the vortex (such as local wind speeds within the vortex). For very large vortices, such as the Great Red Spot, where winds can reach the same order of magnitude as the zonal wind speeds \citep{Wong2021}, this calculation is likely to be incorrect to determine the stability of the resulting vortex. As such, it is important to note that this analysis is targeted at determining the origins of instabilities rather than studying the stability of the vortex itself; we are simply using the location of the vortex as a proxy to diagnose the formation of an instability. 

We use the zonal wind shown in Figure~\ref{fig:zonal_wind} to derive $L_{d,c}$. Given the smoothing profile applied and the sensitivity of the $u_{yy}$ profile to small variations in $u$, we tested different smoothing windows and their corresponding effect on $L_{d,c}^2$. We discuss the sensitivity of the results to the zonal wind profile used in Appendix~\ref{sec:zonal_wind_sensitivity}. For our analysis, we tested both profiles, and found negligible change between the two profiles and the resulting overall distribution of $L_{d,c}$ described below. Due to the higher spatial resolution and the closeness in time of the \citeauthor{Tollefson2017} profile to the vortices studied here, our analysis below uses the OPAL derived wind speeds from 2016.

\begin{figure}
    \centering
    \includegraphics[width=\columnwidth]{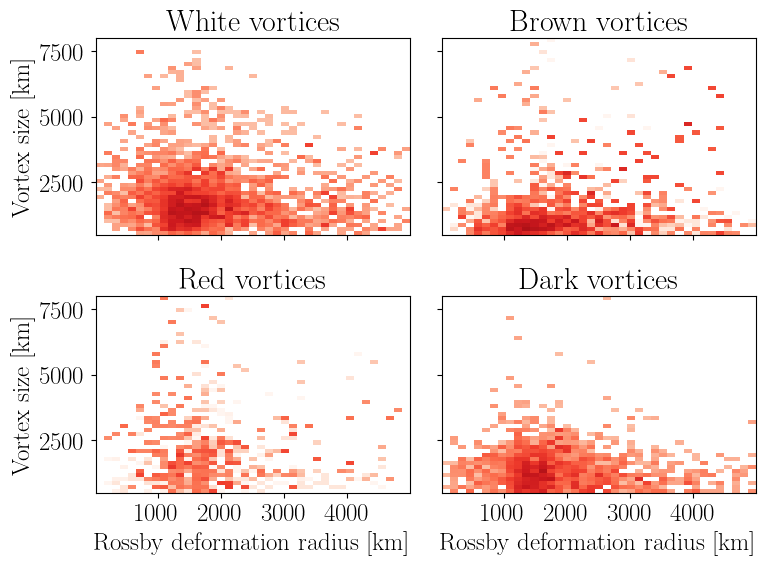}
    \caption{Derived critical Rossby deformation length by vortex color. The histogram is weighted by the inverse of the number of images in each latitudinal bin in order to remove observational bias in the JunoCam data.}
    \label{fig:Ld_size}
\end{figure}

Figure~\ref{fig:Ld_size} shows the relationship between the derived $L_{d, c}$ and corresponding vortex sizes for different categories of vortices. All vortex types show a very poor correlation between the observed size and $L_{d,c}$, except for large white vortices, which are concentrated around $L_{d,c} \sim 1800$ km. However, there is a large distribution that trends to high values of $L_{d,c}$ for all types, with the $98$th percentile of $L_{d,c}$ being about $5000$ km for all vortex types. Since the large range of $L_{d,c}$ occurs only for the smaller vortices, it is possible that the error in calculating $L_{d,c}$ is highly sensitive to small changes in zonal wind properties. Indeed, such smaller vortices predominantly form in high shear regions, and it is likely that we are not able to determine an accurate value of $u_{yy}$ in order to precisely determine the critical $L_{d,c}$ for the vortex. 

\begin{figure}
    \centering
    \includegraphics[width=\columnwidth]{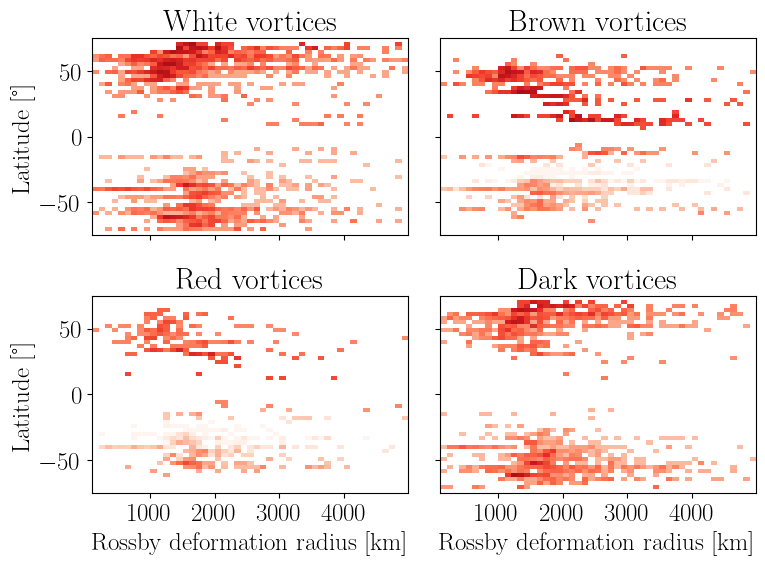}
    \caption{Derived critical Rossby deformation length by vortex color and latitude. There is no observable trend in the mean value with latitude. The histogram is weighted by the inverse of the number of images in each latitudinal bin in order to remove observational bias in the JunoCam data. }
    \label{fig:Ld_lat}
\end{figure}

Figure~\ref{fig:Ld_lat} shows the distribution of $L_{d,c}$ as a function of latitude. There doesn't appear to be a specific trend in the mean value, but the variance increases from the low latitudes towards the poles. This is to be expected since the polar regions have narrow alternating jets, which means that small changes in latitude lead to large variations in zonal wind, and correspondingly large variations in $L_{d,c}$. In these regions, it is difficult to constrain the exact values of $L_{d,c}$, although the peak of the distribution even at these locations corresponds to the global mean value.

\section{Discussion} \label{sec:discussion}

\subsection{Observed vortex properties and vortex classes}
In general, our method provides accurate estimates of vortex properties. Particularly, analysis of volunteer consensus and cluster confidences show that the distribution of observed properties (e.g., size, ellipticity and location) provide high confidence. However, the main confusion in our analysis is the large diversity of vortices across the planet, particularly poleward of $50\degree$. Here, vortices span a large range of sizes (see Figure~\ref{fig:size_lat}), and the mean vortex properties are less accurate than in the mid-latitudes. As such, at these locations, it is perhaps necessary to determine ways to identify and disentangle groups of vortices and derive unique vortex properties to each class separately. 

In this study, the average vortex properties across these large variances still provide meaningful conclusions. For one, we find that each vortex color has unique distributions -- white and dark (dark vortices are interpreted to be cloud free interiors), tend to populate high latitudes, with very few counts near the equator/mid-latitudes. Red and brown vortices (which are likely to be small cyclones and brown barges, respectively) favor the mid-latitudes, and the distribution falls off towards the polar and equatorial regions. 

The red vortices, in particular, do not seem to follow a homogeneous distribution, and instead peak strongly at specific latitudes. From our data, it is difficult to disentangle the role of the fluid dynamical processes with the chemical processes that create specific colors of vortices, but the inhomogeneity of their distributions raises many questions, such as what processes create these vortices, and how they are sustained. Many volunteers have identified small red vortices within the loops of FFRs, so it is likely that the dynamical conditions inside FFR loops are conducive of these types of vortices, but a comprehensive examination of the local fluid dynamics is required to understand the uniqueness of these ovals. Do these red vortices require a specific chromophore to be present? Are there locations where the fluid dynamics present the opportunity for these vortices to exist, but the chromophore is absent, or is the chromophore globally abundant?
These questions are difficult to answer through observations from \JunoCam{} alone, and possibly require a much more thorough analysis involving the characterization of aerosols within these vortices through atmospheric retrieval analysis, and the use of numerical fluid dynamical models to constrain their dynamical origins. 

Furthermore, the footprint of the JunoCam instrument raises a few issues worth mentioning. Specifically, all the JunoCam images have a wide longitudinal footprint near the poles and a very narrow viewing geometry near the lower latitudes (when the spacecraft is closer to the planet). While we have accounted for this in our analysis here by weighting the distributions by the inverse of the number of images shown to volunteers at each latitude bin, it is nevertheless impossible to accurately draw conclusions about the distribution of vortices in the lower latitudes. One avenue is to study this region with our datasets such as Hubble or ground-based images which have better coverage at these latitudes.

\subsection{Log normality of vortex sizes and vortex stability}
One of the interesting insights into the size distribution of vortices on Jupiter show that they are a good fit to log-normal distributions. Such distributions are generally a result of a balance between growth and dissipation processes \citep[see, e.g., growth of rain droplets, ][]{MarshallPalmer1948}, where the growth rate scales inversely with size and dissipation rate scales proportionally. Therefore, the mean of the distribution provides information about the growth rates associated with both processes. In turbulence, log-normality is due to stochastic processes that combine multiplicatively and transfer energy across different scales \citep{Mouri2009}. Indeed, turbulent kinetic energies and eddy sizes observed in Earth oceanic eddies generally follow a log-normal distribution \citep{Tang2019}. Studies of oceanic eddies have shown amazing similarity to the Jovian atmosphere, particularly in the avenue of instabilities \citep{Siegelman2022}.

In our case, the vortex size distributions provide information about the length scales at which vortices form and dissipate. As expected, we find that the mean vortex size across all colors is about 1500-2000 km, which is consistent with derivations of the mean Rossby deformation radius ($L_d$) on Jupiter, and the expected size of eddies on the planet \citep{Vasavada2005}. However, there are exceptions to this: particularly, we see that several long-lived vortices, such as the Great Red Spot, Oval BA and the white ovals at $40\degree$ S latitude, are much larger than this derived $L_d$, and seem to be stable. There are also examples of brown barges expanding to large radii and destabilizing to result in FFRs \citep{Hueso2022,Inurrigarro2022}. 

Therefore, vortex stability is likely a function of unobserved vortex properties, such as their vertical thickness, local atmospheric stability, and possibly even the types of processes that create them (e.g., convective processes vs dynamical shear). The statistics defined here, particularly given the strong relation to the derived $L_d$, allude to the fact that most of the vortices observed on Jupiter are destabilized at sizes larger than the atmospheric $L_d$. This rule is most relevant to shallow eddies, implying that most of the vortices observed are likely not vertically extended. These parameters, however, will need additional observations to identify longevity of these vortices (particularly those at the cusp between stability and instability) to constrain their 3-dimensional structure. From our analysis, we see that the zonal wind profile shows some correlation with vortex properties for certain types of vortices, but a `one-size-fits-all' type of theory is incomplete with just the 1-dimensional information provided by the cloud-top zonal wind speeds. Indeed, information about the vertical structure of the atmosphere (e.g., static stability, vertical wind shear, etc.) would provide better information on how these ovals form on Jupiter, since it would specifically allow us to calculate both the value and gradients of potential vorticity, which have stronger effects on vortex evolution \citep{Dowling1995Review,LeBeau1998,Read2006}.

\subsection{Critical Rossby deformation length}
The critical Rossby deformation length derived here follows the assumption that the environment surrounding the vortex is representative of the environment that generated the vortices. While this may not be true for locations with strong meridional shear, or even long-lived vortices (such as the Great Red spot, or the Oval BA), the statistics provided by the large number of observed (likely transient) vortices, provides a good estimate, and outweighs the small fraction of long-lived ovals. As such, the outcomes of the derivation of these properties provides valuable insight into the jovian atmosphere. 

First, there is very little variation in $L_d$ with latitude, with most vortices showing a mean $L_{d,c}$ of around 1800 km. As stated above, while the mean remains the same, the variance of $L_{d,c}$ increases towards the poles as vortex locations and sizes become much less homogeneous. It is possible that due to the turbulence of the atmosphere, and increased concentration of turbulent eddies towards the poles, there does exist large variations in the atmospheric $L_d$. This is likely due to both the thickness of the fluid layer and the static stability being strongly affected by vorticity \citep{DowlingGierasch1989}, and therefore a large diversity in vortices at these locations lead to a large variation in $L_d$. To further constrain the distribution of $L_d$ here, we require observations of the 2D wind structure, and include baroclinic effects in the stability criteria that we have derived here. Furthermore, we must also constrain vortex stability and their temporal evolution within these locations in order to ensure that our assumptions above are valid. This will require both accurate observations of the 3D structure of the atmosphere (to determine both the 2D dynamics and the vertical stratifications of the atmosphere) and the use of numerical models that accurately treat convection. 

\section{Conclusions and Future Work} \label{sec:conclusions}
In summary, we built a citizen science project, the Jovian Vortex Hunter, on Zooniverse, to derive and study the distribution of vortices, and their associated properties on Jupiter. We presented crops from \JunoCam{} images to citizen scientists on Zooniverse, and asked them to identify and annotate vortices by color. Our resulting dataset contains over 7000 vortices across perijoves 13 through 36 (i.e., from 2018 through 2021), and we have derived distributions of vortex sizes, aspect ratios and locations across the planet. We find that there is a strong correlation between the cloud chemistry (i.e., color of the vortex) and the associated vortex properties -- white and brown vortices are the largest, while red and cloud-free (dark) vortices are the smallest. White and dark vortices are preferentially found near the poles, brown vortices are most prominent in the mid-latitudes, while red vortices form at unique latitudes non-homogeneously. 

Furthermore, we have used a simplification of Arnol'd's atmospheric stability criterion to derive a minimum value for the Rossby deformation radius of Jupiter. We find that this value is roughly $1800$ km, and is generally consistent across the planet.  This mean $L_{d,c} \sim 1800$ km for all vortex types is consistent with previous studies of Jovian vortex dynamics \citep{Cho2001,Showman2007} and measurements of vertical stratification \citep{Magalhaes2002}, although recent measurements from JIRAM suggest a lower value closer to 1000 km \citep{Moriconi2020}. However, we do note that there are several considerations which must be validated in order to improve the accuracy of this study. First, since \JunoCam{} data does not have a large temporal coverage, it is difficult to interpret vortex stability. As such, it is impossible to know which of the vortices studied here are stable, and which are transient. It would be interesting to extend this study to later perijoves and correlate vortices across large temporal scales to study their morphological evolution. Second, our derivation of $L_{d,c}$ uses only the zonal wind information and neglects the in-vortex wind speeds. To derive a holistic and self-consistent theory of vortex dynamics on Jupiter, we require better constraints on these properties. This will enable us to generalize the lessons learnt here to infer global properties of fluid dynamics on Jupiter, and other Gas Giant planets. Finally, this study covers roughly 3 years worth of imaging data to derive the catalog, but there are still several perijoves worth of images that could be used to increase the size of the catalog, which would be useful to derive long term trends in the distributions derived here. We intend to continue this project and feature these remaining perijoves, but are also investigating methods to extend the project outside of just the JunoCam data. For one, we believe it would be interesting to look at historical data (specifically during periods before/after intense convective activity, such as the South Equatorial Belt and North Tropical Belt outbreaks) to study the change in the vortex properties during these epochs. Furthermore, the improvements in computing (specifically machine- and deep-learning techniques) hold promise in transferring the ability to detect vortices in JunoCam data to other datasets (e.g., OPAL or amateur observation), where the volume of images are equally large. We intend to investigate these avenues, which we expect hold promising insights into the jovian atmosphere.

\begin{acknowledgments}
We would like to thank all the volunteers on the Jovian Vortex Hunter project for their incredible contribution to this study. This study would not be possible without their efforts. RS, KM and LF were funded in part by NASA Award \#80NSSC20M0057. RS was also partly funded by NASA Award \#80NSSC22K0804.
\end{acknowledgments}



\appendix

\section{Negative $L_{d,c}^2$ and sensitivity to zonal wind profile}
\label{sec:zonal_wind_sensitivity}
In the form presented in Equation~\ref{eq:ldc}, it is possible to obtain a negative value of $L_{d,c}^2$, which is unphysical. We attribute these values to three issues in our analysis: firstly, the simplification made to the equation neglects the effect of vertical wind shear and baroclinicity in the derivation of $``Ma"$, making the derivation unsuitable for locations with large vorticity or steep wind shear. Secondly, small variations (numerical or physical) in the zonal wind propagate to large variations in the calculation of $u_{yy}$, producing peaks in $u_{yy}$ which overshoot the $\beta$ curve (see Figure~\ref{fig:zonal_wind}). Thirdly, poleward of $\sim 50\degree$ or near zero-points in the shear (e.g., at $40\degree$ S), small variations in the zonal wind value can produce negative values of $L_{d,c}^2$. Given these issues, we detail our sensitivity tests below to understand the  effects (possibly numerical) of our analysis on the resulting $L_{d,c}^2$ profile in an effort to constrain the precision of our derived $L_{d,c}$ profile.

\begin{figure}
    \centering
    \includegraphics[width=\columnwidth]{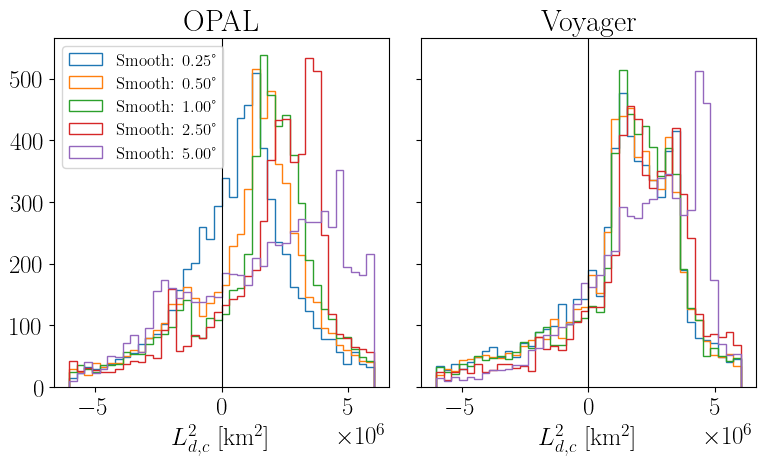}
    \caption{Distribution of $L_{d,c}^2$ as a function of the smooth window size for the two zonal wind profiles used.}
    \label{fig:ldc_smooth}
\end{figure}

Given that the zonal wind profiles are derived from cloud tracking methods, there is significant noise in the sub-degree resolution, which is compounded when taking the second derivative to calculate $u_{yy}$. To remove this noise, we applied a moving average smoothing to our profiles. We tested different values of smoothing window sizes from $0.25\degree$ to $5\degree$, as shown in Figure~\ref{fig:ldc_smooth}. For intermediate values between $0.5$-$1\degree$, we found very little difference in the distribution, while low smoothing produced a shift to the left (specifically for the OPAL data, which has a resolution of $0.05\degree$), while high smoothing flattened the distribution and produced a shift to the right. We find that the $1\degree$ resolution provides the most consistent result between the two profiles and offers a balance between the noise generated by the smaller smoothing windows and the flattening created by the larger windows. In our nominal analysis, we applied a moving average of $1\degree$ to both the Voyager and OPAL profiles, which is larger than the smoothing applied to the OPAL profile \citep{Tollefson2017} but slightly smaller than other analyses that derive $u_{yy}$ \citep[e.g.,][]{Read2006}.

\begin{figure}
    \centering
    \includegraphics[width=\columnwidth]{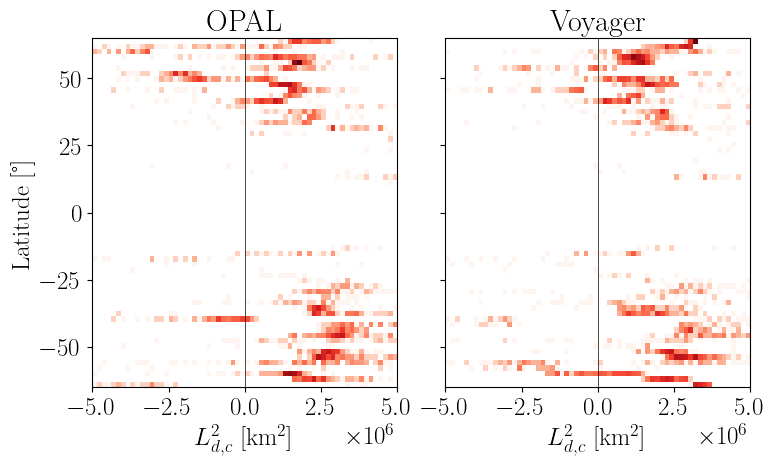}
    \caption{Distribution of $L_{d,c}^2$ with latitude for the two zonal wind profiles used.}
    \label{fig:ldc_lat}
\end{figure}

Figure~\ref{fig:ldc_lat} shows the distribution of $L_{d,c}^2$ as a function of latitude for all the vortices in the catalog. We find that most of the negative values of $L_{d,c}^2$ appear closer to the poles, where either the velocity changes signs or $\beta$ is small such that it is comparable to the errors in the $u_{yy}$ curve. The exceptions to this are two locations in the OPAL data at $\sim15\degree$ S and $\sim 40\degree$ S. 

For the former case, we find that this is due to the peak in the $u_{yy}$ profile that results from a sharp increase in the OPAL zonal wind speed over the Voyager speed at $\sim 12\degree$ S. The 2016 zonal wind at this latitude is consistently higher than other profiles derived in \citeauthor{Tollefson2017}, and indeed appears to be a non-negligible increase over the uncertainty. It is unclear whether this increase in wind speed is driven by (or drives) convective activity \citep{Wong2020} or an artifact of observing different pressures \citep{Wong2021}, and requires further study.

In the latter case, at $40\degree$ S, this is due to the OPAL zonal wind showing a weak westward component at $40\degree$ S. However, given that the uncertainty in the derived wind speed is $\sim \pm 20$ m/s, it is possible that the negative $L_{d,c}^2$ is not a physical characteristic, but rather within the uncertainty in the wind speeds. Given the presence of longitudinal variability (e.g., vortices, FFRs, etc.), it is difficult to interpret whether these zonally averaged profiles are truly indicative of the local wind speed at this latitude.

In summary, while we do observe negative $L_{d,c}^2$ in our analysis, it is more than likely that they are numerical effects rather than physical ones. However, constraining the uncertainty in the value of $L_{d,c}^2$ at these locations is nearly impossible due to the simplifications made to the derivation of the instability criterion and the compounding errors from the gradients in the zonal wind profiles. For the remainder of the analysis, we proceed with the positive values of $L_{d,c}^2$ which still show a remarkably strong peak at $\sim 2.5\times10^6$ km$^2$ across all the zonal wind profiles used.  


\bibliography{ref}{}
\bibliographystyle{aasjournal}



\end{document}